\documentclass[author-year, prd, amsmath, amssymb, longbibliography, floatfix, reprint, superscriptaddress, a4]{revtex4-1}

\usepackage{bm}
\usepackage[toc, page]{appendix}
\usepackage[alsoload=hep]{siunitx}
\usepackage{tikzscale}
\usepackage{pgfplotstable}
\usepackage[mode=image, extension=.pdf]{standalone}
\usepackage{bigints}
\usepackage{booktabs}
\usepackage{ifthen}
\usepackage{tensor}
\usepackage{xfrac}
\usepackage{parskip}
\usepackage[
	top    = 1.95cm,
	bottom = 1.9cm,
	left   = 1.5cm,
	right  = 1.5cm]
{geometry}

\definecolor{color1}{RGB}{202,0,32}
\definecolor{color2}{RGB}{244,165,130}
\definecolor{color3}{RGB}{146,197,222}
\definecolor{color4}{RGB}{5,113,176}

\hypersetup{
	colorlinks,
	linkcolor={color4},
	citecolor={color4},
	urlcolor={color4}
}

\AtBeginDocument{
	\heavyrulewidth=.08em
	\lightrulewidth=.05em
	\cmidrulewidth=.03em
	\belowrulesep=.65ex
	\belowbottomsep=0pt
	\aboverulesep=.4ex
	\abovetopsep=0pt
	\cmidrulesep=\doublerulesep
	\cmidrulekern=.5em
	\defaultaddspace=.5em
}

\newcommand{\vect}[1]{\bm{\mathrm{#1}}}

\newcommand{\rmi}{\mathrm{i} \mkern1mu} 
\newcommand{\dd}{\textrm{d}}
\newcommand{\np}[1]{^{\!\!#1}}
\DeclareMathOperator{\arcsec}{arcsec}

\newcommand{\G}{\Gamma}
\newcommand{\e}{\epsilon}
\newcommand{\s}[1][]{%
	\ifthenelse{\equal{#1}{}}{S(\Theta)}{S^{#1}(\Theta)}%
}
\newcommand{\ssqrt}{\sqrt{\e (2-\e) + (1-\e)^{2}\s[2]}}
\newcommand{\eln}{\ln\!\left[\frac{2 - \e}{\e}\right]}

\newcommand{\sln}{\ln\!\left[\frac{\ssqrt + (1-\e)\s}{\sqrt{\e (2-\e)}}\right]}

\newcommand{\lns}{\ln\!\left[\s\right]}

\newcommand*\pFqskip{8mu}
\catcode`,\active
\newcommand*\pFq{\begingroup
        \catcode`\,\active
        \def ,{\mskip\pFqskip\relax}%
        \dopFq
}
\catcode`\,12
\def\dopFq#1#2#3#4#5{%
        {}_{#1}F_{#2}\biggl(\displaystyle {\genfrac{}{}{0pt}{0}{#3}{#4}};#5\biggr)%
        \endgroup
}

\ExplSyntaxOn
\DeclareExpandableDocumentCommand{\RepQuad}{m}
{\int_compare:nT { #1 > 0 }
	{\quad \prg_replicate:nn { #1 - 1 } {\quad}}}
\ExplSyntaxOff

\newcommand{\mathsc}[1]{\mbox{\tiny{\(#1\)}}}

\renewcommand{\thetable}{\arabic{table}}

\makeatletter 
	\renewcommand{\fnum@figure}{Fig.~\thefigure}
\makeatother

\makeatletter 
	\renewcommand{\fnum@table}{Table~\thetable}
\makeatother

\allowdisplaybreaks[4]

\def\apjl{Astrophys.~J.~Lett.}
\def\prd{Phys.~Rev.~D}
\def\prl{Phys.~Rev.~Lett.}

\def\spaw{Sitzungsber. Preuss. Akad. Wiss. Berlin (Math. Phys.)}
\def\jfi{J.~Frankl.~Inst.}

\begin{document}


\title[Astrometric Effects of Gravitational Wave Backgrounds with non-Luminal Propagation Speeds]{Astrometric Effects of Gravitational Wave Backgrounds \texorpdfstring{\\}{} with non-Luminal Propagation Speeds}

\author{Deyan P. Mihaylov}
\email{deyan@aei.mpg.de}
\affiliation{Institute of Astronomy, University of Cambridge, Madingley Road, Cambridge, CB3 0HA, UK}
\affiliation{School of Mathematics, University of Edinburgh,~Peter Guthrie Tait Road, Edinburgh, EH9 3FD, UK}
\affiliation{Max Planck Institute for Gravitational Physics (Albert Einstein Institute), Potsdam Science Park, Potsdam, D-14476, Germany}

\author{Christopher J. Moore}
\email{cmoore@star.sr.bham.ac.uk}
\affiliation{Institute for Gravitational Wave Astronomy and School of Physics and Astronomy, University of Birmingham, Edgbaston, Birmingham B15 2TT, UK}

\author{Jonathan~R.~Gair}
\email{jonathan.gair@aei.mpg.de}
\affiliation{School of Mathematics, University of Edinburgh,~Peter Guthrie Tait Road, Edinburgh, EH9 3FD, UK}
\affiliation{Max Planck Institute for Gravitational Physics (Albert Einstein Institute), Potsdam Science Park, Potsdam, D-14476, Germany}

\author{Anthony Lasenby}
\email{a.n.lasenby@mrao.cam.ac.uk}
\affiliation{Astrophysics Group, Cavendish Laboratory, J J Thomson Avenue, Cambridge CB3 0HE, UK}
\affiliation{Kavli Institute for Cosmology, Madingley Road, Cambridge CB3 0HA, UK}

\author{Gerard Gilmore}
\email{gil@ast.cam.ac.uk}
\affiliation{Institute of Astronomy, University of Cambridge, Madingley Road, Cambridge, CB3 0HA, UK}

\date{\today}

\begin{abstract}
A passing gravitational wave causes a deflection in the apparent astrometric positions of distant stars. The effect of the speed of the gravitational wave on this astrometric shift is discussed. A stochastic background of gravitational waves would result in a pattern of astrometric deflections which are correlated on large angular scales. These correlations are quantified and investigated for backgrounds of gravitational waves with sub- and super-luminal group velocities. The statistical properties of the correlations are depicted in two equivalent and related ways: as correlation curves and as angular power spectra. Sub-(super-)luminal gravitational wave backgrounds have the effect of enhancing (suppressing) the power in low-order angular modes. Analytical representations of the redshift-redshift and redshift-astrometry correlations are also derived. The potential for using this effect for constraining the speed of gravity is discussed.
\end{abstract}

\maketitle

\setlength{\parskip}{10pt}
\setlength{\parindent}{10pt}

\section{Introduction}
\noindent
In his seminal papers on gravitational waves (\textsc{gw}s), Einstein demonstrated that the speed of those waves in the theory of general relativity is equal to the speed of light \citep{1916SPAW.......688E, 1918SPAW154E, 1937FrInJ.223...43E}. The near-simultaneous observation of the \textsc{gw} signal GW170817 in the frequency band \(20-2000\,\mathrm{Hz}\) and the gamma-ray burst GRB 170817A provided a spectacular confirmation of this prediction \cite{2017PhRvL.119p1101A, 2041-8205-848-2-L12, 2017ApJ...848L..13A, 2017ApJ...848L..13A}. This article considers the astrometric effects of a non-luminal propagation speed for stochastic backgrounds of much lower frequency \textsc{gw}s.

Investigations into the effects of \textsc{gw}s on the observed properties of light have a long history. For example, the periodic change in the brightness of a light source -- \textsc{gw}-induced scintillation -- was considered as early as 1966 by \citep{PhysRev.142.825}. \textsc{gw}s also cause astrometric changes in the positions of distant objects; the dominant effect is a distortion of the sky due to the local metric perturbation around the observer, although there is a secondary effect due to the metric perturbation at the light source \cite{1996ApJ...465..566P} (see also \cite{PhysRevD.83.024024, 2018PhRvD..97l4058M}). Low frequency \textsc{gw}s induce apparent proper motions in distant objects; this effect has been used previously to constrain the energy in low frequency \textsc{gw}s by observing quasar proper motions \citep{Gwinn:1996gv,Darling:2018hmc}, but can also be used in the future with \emph{Gaia} data. The advantage of \emph{Gaia} is that it retains its sensitivity at higher frequencies too (up to \(\sim \SI{1e-6}{\hertz}\)) \citep{PhysRevLett.119.261102, Klioner:2017asb}.

\textsc{gw}s in Einstein's theory of general relativity have only two polarization states which travel at the speed of light. Modified theories of gravity may contain up to six polarizations travelling at a range of speeds \citep{Will:2014kxa}. The authors have previously explored the astrometric effects of a stochastic \textsc{gw} background with non-Einsteinian polarizations \citep{2018PhRvD..97l4058M}; the background produces a stochastic vector field of astrometric deflections correlated across the sky. This effect was also studied by \citep{2018PhRvD..98b4020O} who also considered decomposing the astrometric deflection in terms of vector spherical harmonics (\textsc{vsh}). This article studies these cross-correlations, both at the level of correlations between the components of the astrometric deflection vector field and correlations between the components in its \textsc{vsh} decomposition, for \textsc{gw}s with arbitrary polarization and propagating at sub- or super-luminal group velocities.

While within \textsc{gr} gravitational waves move at the speed of light, modified theories of gravity permit endowing the \textsc{gw} degrees of freedom with a mass, in which case they would propagate at a speed lower than the speed of light. The group velocity of the \textsc{gw} \(v_{\textsc{gw}}\) may be parametrised by \(\e \in [-1, 1]\) according to \(\e = 1 - v_{\textsc{gw}} / c\). There are currently no experimental constraints on the propagation of low frequency \textsc{gw}s, i.e. \(f<\SI{e-6}{\hertz}\), where \emph{Gaia} and pulsar timing arrays are sensitive \citep{PhysRevD.83.024024, Manchester:2015mda}. \textsc{ligo}/Virgo observations have placed constraints at higher frequencies and verified that the propagation of the observed \textsc{gw}s in the range \(\SIrange{e1}{e3}{\hertz}\) is in accordance with the predictions of \textsc{gr}. As an example of the type of test possible using the \textsc{ligo}/Virgo observations, consider the possibility that gravitons are dispersed in vacuum like massive particles; in this case the speed parameter \(\e>0\) is a function of both the graviton mass and the \textsc{gw} frequency,
\begin{align}\label{eq:edef}
\e = 1 - \sqrt{1-\frac{m_{\mathrm{g}}^{2} c^{4}}{\hbar^{2}\omega^{2}}}\,.
\end{align}
Under this assumption, the combined \textsc{ligo}/Virgo observations of GW15914, GW151226, and GW170104 constrain the graviton mass to be \(m_{\mathrm{g}} \leq \SI[per-mode=symbol]{7.7e-23}{\eV\per\clight\squared}\) \citep{2017PhRvL.118v1101A} which gives the impressively tight constraint \(\e \lesssim \SI{e-20}{}\) at \(f \approx \SI{100}{\hertz}\). Continuing to assume that \textsc{gw}s are dispersed in vacuum like massive particles, and using eq.~(\ref{eq:edef}) to extrapolate this constraint across \(\sim 10\) orders of magnitude in frequency gives the weakest possible constraint of \(\e \lesssim 1\) at a frequency of \(f \approx \SI{e-8}{\hertz}\) (no sensible constraint is possible at lower frequencies, though). The scaling of \(\e\) with the \textsc{gw} frequency in eq.~(\ref{eq:edef}) means that \(\e\) is only weakly constrained (or not constrained at all) across the frequency range of interest for pulsar timing and astrometric observations; consequently, any new bound that these techniques are able to place will likely represent an improvement.

\begin{figure}[b!]
\includegraphics[scale=1]{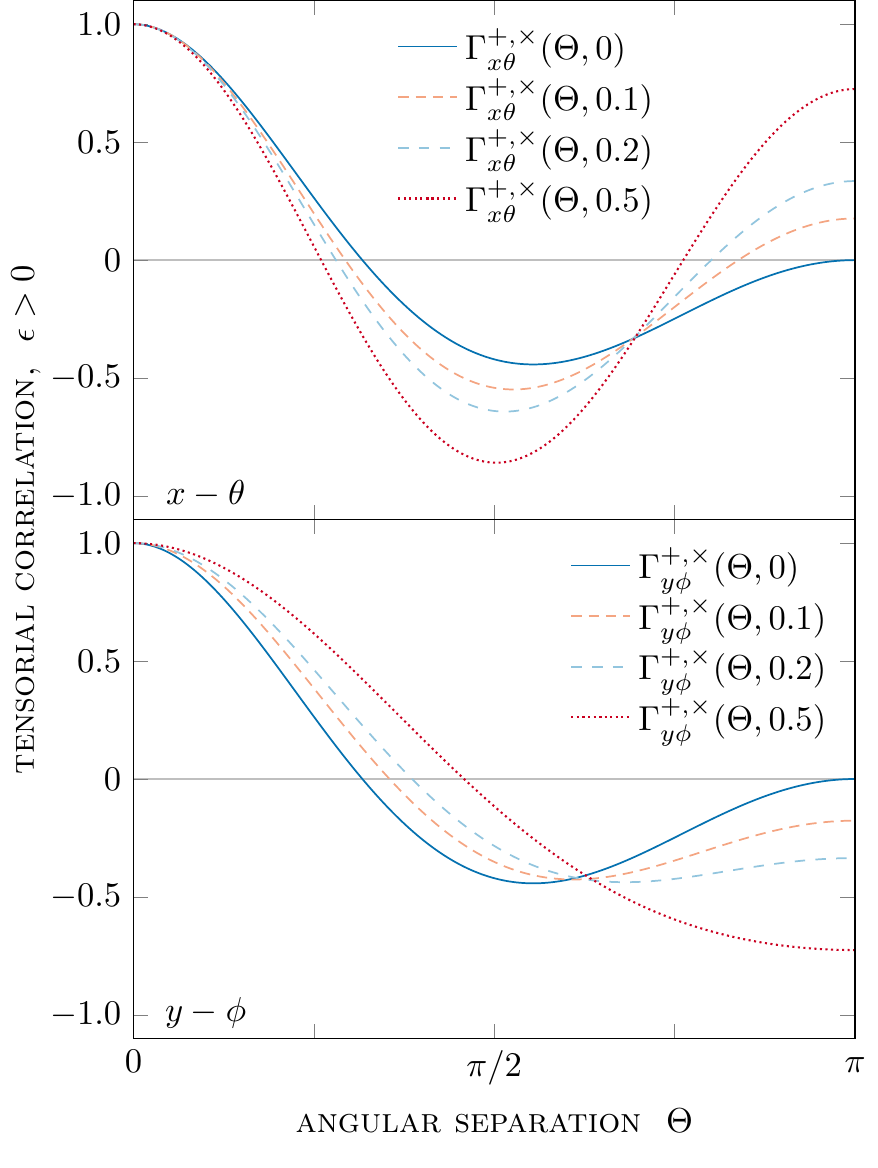}
\caption{The \(\G_{x\theta}^{+, \times}\) and \(\G_{y\phi}^{+, \times}\) cross-correlation functions for a background of sub-luminal (\(\e > 0\)) tensorial transverse-traceless \textsc{gw}s for 4 different values of \(\e \in \{0, 0.1, 0.2, 0.5\}\) using the distant-source limit of the astrometric response. While the \(x-\theta\) and \(y-\phi\) curves are the same in the \(\e = 0\) case, the degeneracy breaks down for any \(\e > 0\). The \(\e = 0\) case is predominantly quadrupole (\(\ell = 2\)) in the sense that the curves have two zeros in the range shown; the curve is not a pure quadrupole, see Fig.~\ref{fig:subluminalTensorialCoeffs}.}
\label{fig:subluminalTensorialCorr}
\end{figure}

\begin{figure}[b!]
\includegraphics[scale=1.0]{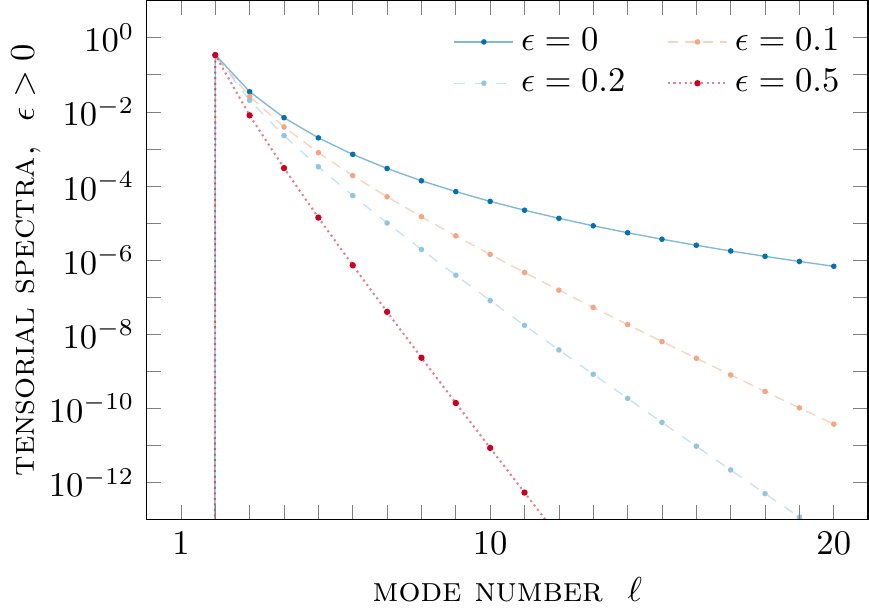}
\caption{The \(C_{\ell}^{+, \times}\) angular power spectra for a background of sub-luminal (\(\e > 0\)) tensorial transverse-traceless \textsc{gw}s for 4 different values of \(\e \in \{0, 0.1, 0.2, 0.5\}\), corresponding to each of the curves in Fig.~\ref{fig:subluminalTensorialCorr}. A higher value of \(\e\) has the effect of suppressing the higher multipole moments thus making the resultant pattern increasingly dominated by the quadrupole (\(\ell=2\)).}
\label{fig:subluminalTensorialCoeffs}
\end{figure}

In addition to sub-luminal group velocities (\(\e>0\)), modified theories of gravity also predict additional \textsc{gw} polarization states to those present in \textsc{gr}. For a massive graviton, with spin \(s=2\) there are \(2s+1=5\) different polarization states available (as stated above, the most general theories have 6 polarization states; however, massive gravity has one less, as the two scalar modes are indistinguishable, see \citep{Fierz:1939ix, Comelli:2012vz, Boulware1972227}), versus just the two transverse polarization states in \textsc{gr}. A recent review of such theories \citep{Cardoso:2018zhm} demonstrates how these modes correspond to unusual emission mechanisms (monopole and dipole) and emphasises that the best range for measuring these effects from real objects will be the ultra-low frequencies \(f \approx \SI{e-9}{\hertz}\), which is well suited to pulsar timing and astrometric observations. Additional polarization modes occur in other modified gravity theories as well, such as the additional longitudinal mode found in some \(f(R)\) theories \citep{Capozziello:2008rq}. For these reasons, this article considers non-luminal \textsc{gw} group velocities in the full set of non-Einsteinian polarizations, as discussed e.g. in \citep{2018PhRvD..97l4058M}.

The above discussion assumed that gravitons disperse as massive particles and it was this assumption that allowed the \textsc{ligo}/Virgo constraint to be extrapolated across many orders of magnitude. This is an extremely restrictive assumption seeing as there are a plethora of plausible theoretical reasons (besides a simple graviton mass) why \textsc{gw}s may propagate at non-luminal group velocities (e.g. in theories violating Lorentz invariance such as Einstein-aether \citep{Jacobson:2000xp, 2004PhRvD..70b4003J} or chronometric \citep{Blas:2010hb} theories). Any of these alternative possibilities would invalidate the above extrapolation. Therefore the above constraints on the parameter \(\e\) at low frequencies should be treated with a large degree of caution. The conservative point of view, and the view adopted in this paper, is to consider the speed parameter \(\e\) as currently being effectively unconstrained in the frequency band of interest and to view pulsar timing or astrometry as providing a test of \textsc{gr} in a new dynamical regime with a much longer timescale.

\begin{figure}[t]
\includegraphics[scale=1]{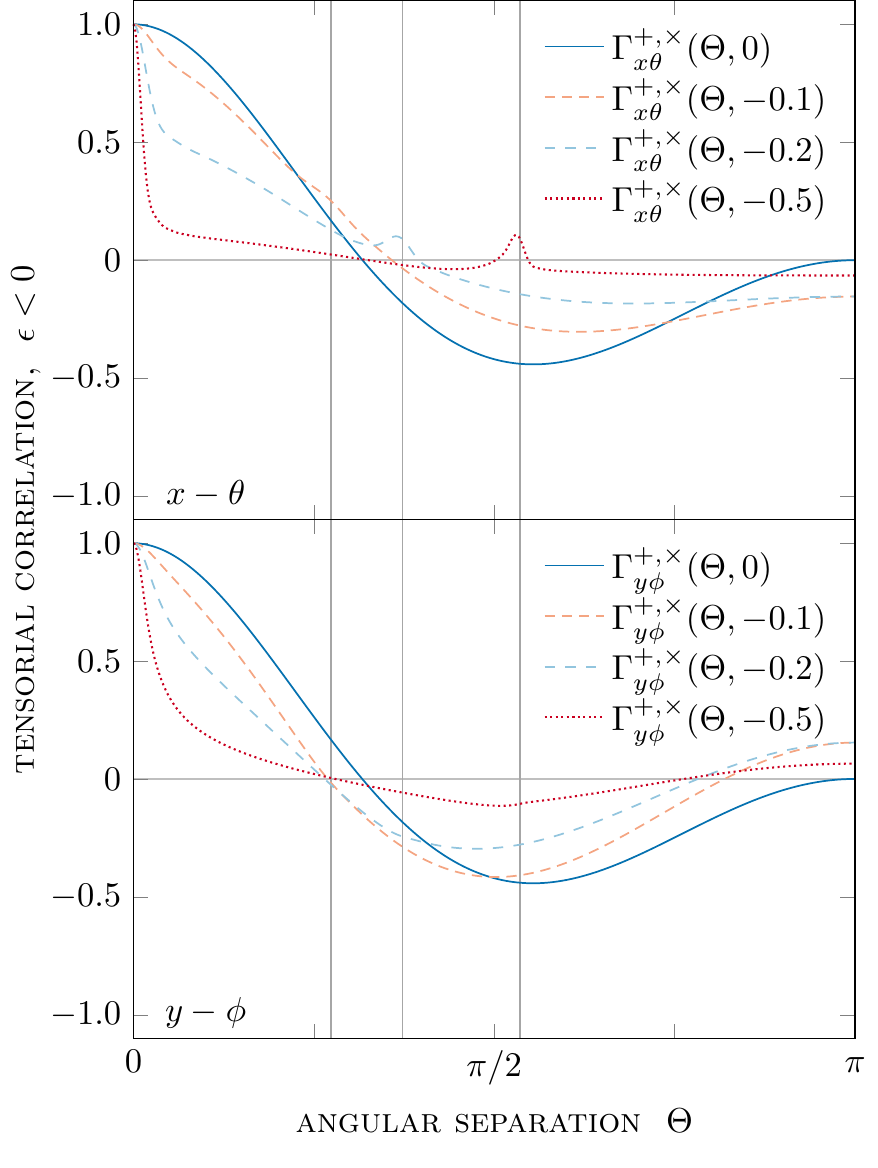}
\caption{The \(\G_{x\theta}^{+, \times}\) and \(\G_{y\phi}^{+, \times}\) cross-correlation functions for a background of super-luminal (\(\e\!<\!0\)) tensorial transverse-traceless \textsc{gw}s for 4 different values of \(\e \in \{0, -0.1, -0.2, -0.5\}\) using the astrometric response with \(d = 100\). The vertical lines mark the angles of the cones on which the astrometric response would be divergent if the distant-source limit formula was used. See Fig.~\ref{fig:superluminalTensorialCoeffs} for the spectral composition of these curves.}
\label{fig:superluminalTensorialCorr}
\end{figure}

\begin{figure}[t]
\includegraphics[scale=1]{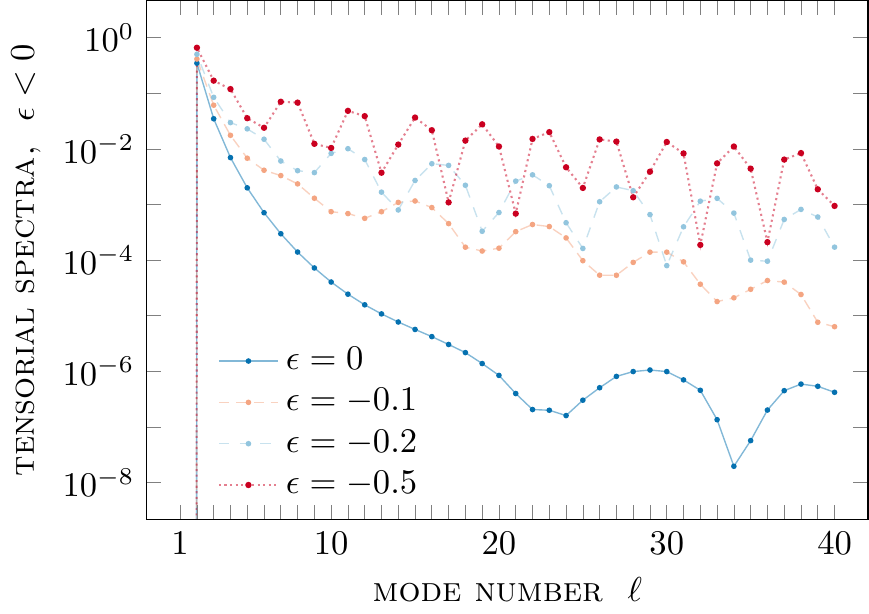}
\caption{The \(C_{\ell}^{+, \times}\) angular power spectra for a background of super-luminal (\(\e < 0\)) tensorial transverse-traceless \textsc{gw}s for 4 different values of \(\e \in \{0, -0.1, -0.2, -0.5\}\). A lower value of \(\e\) has the effect of increasing the power at higher multipoles. The oscillations which are visible in all of the spectra are due to the phases of the star terms in the astrometric shift formula.}
\label{fig:superluminalTensorialCoeffs}
\end{figure}

A mismatch between the speed of the \textsc{gw} and the speed of the photons would incur a qualitative difference in the astrometric response. For the correlated astrometric pattern due to a stochastic background, this will translate into a different pattern of correlated astrometric deflections over the sky. Therefore, astrometric observations in principle allow the speed of \textsc{gw}s to be constrained from both sides, both sub- and super-luminal, and could provide an important, relatively model independent test of general relativity at low frequencies.

Ever since the general theory of relativity was established, there have been attempts to measure the speed of gravity and deviations from the predicted value. The first high-accuracy method which was used involved binary pulsars (like the one studied by Hulse and Taylor). The orbits of binary pulsars decay due to emission of energy as gravitational radiation, and the rate of this decay depends on the speed of gravity. This measurement indirectly confirmed that this speed is within 1\% of the speed of light \citep{Will:2014kxa} in the vicinity of \SI{e2}{\hertz} (where \textsc{ligo} operates too).

Currently, the most stringent limits on the speed of \textsc{gw}s come from the distant GW170104 binary black hole merger, with variation in the \textsc{gw} speed to below a few parts in \(10^{15}\) \citep{2017PhRvL.118v1101A}. Such recent attempts to measure the speed of gravity have, however, dealt with the effect at high \textsc{gw} frequency only, and in the case of binary neutron star mergers, have depended on a particular model for the time delay between \textsc{gw} and gamma-ray burst emission; the speed of gravity remains to be tested in the low-frequency regime, and in a model-independent manner.

Recently, ultra-precise astrometric measurements from the \emph{Gaia} mission \citep{2016A&A...595A...1G, 2016A&A...595A...2G, 2018A&A...616A...1G} have renewed interest in pursuing astrometric detection of \textsc{gw}s \citep{Darling:2018hmc, 1990NCimB.105.1141B, Gwinn:1996gv, PhysRevD.83.024024, 2012A&A...547A..59M, Klioner:2017asb, PhysRevLett.119.261102}. This approach allows probing for both individual monochromatic events, as well as traces of stochastic \textsc{gw} backgrounds. Backgrounds of astrophysical or cosmological origin can be understood as the result of the superposition of a large number of uncorrelated and random individual events (e.g. binary mergers). A stationary, isotropic, unpolarized, Gaussian stochastic background of \textsc{gw}s would result in a stochastic pattern of astrometric measurements on the sky. In \citep{2018PhRvD..97l4058M} the authors have shown that the correlation between the astrometric response at two points is completely specified by two functions which depend solely on the angular separation of these two points.

Previous studies have investigated the possibility of using pulsar timing array (\textsc{pta}) observations to constrain the speed of \textsc{gw}s (and the graviton mass) \citep{Baskaran:2008za, PhysRevD.78.089901, Lee:2014awa}. \textsc{pta}s are sensitive to \textsc{gw}s in a similar low frequency bandwidth to \emph{Gaia}, and it has been predicted that \(0.4\%\) limit on the deviation of the speed of \textsc{gw}s from the speed of light may be attainable \citep{Baskaran:2008za, PhysRevD.78.089901}. In practice, pulsar timing arrays and astrometry measurements may complement each other by independently providing evidence for a \textsc{gw} background at low frequencies. In addition, it might be possible to cross-correlate astrometric and redshift effects to further accelerate and verify detection prospects \citep{2018PhRvD..97l4058M}. The effect of a non-luminal \textsc{gw} group velocity on these methods is also considered in this article.

\begin{figure}[b]
\includegraphics[scale=1]{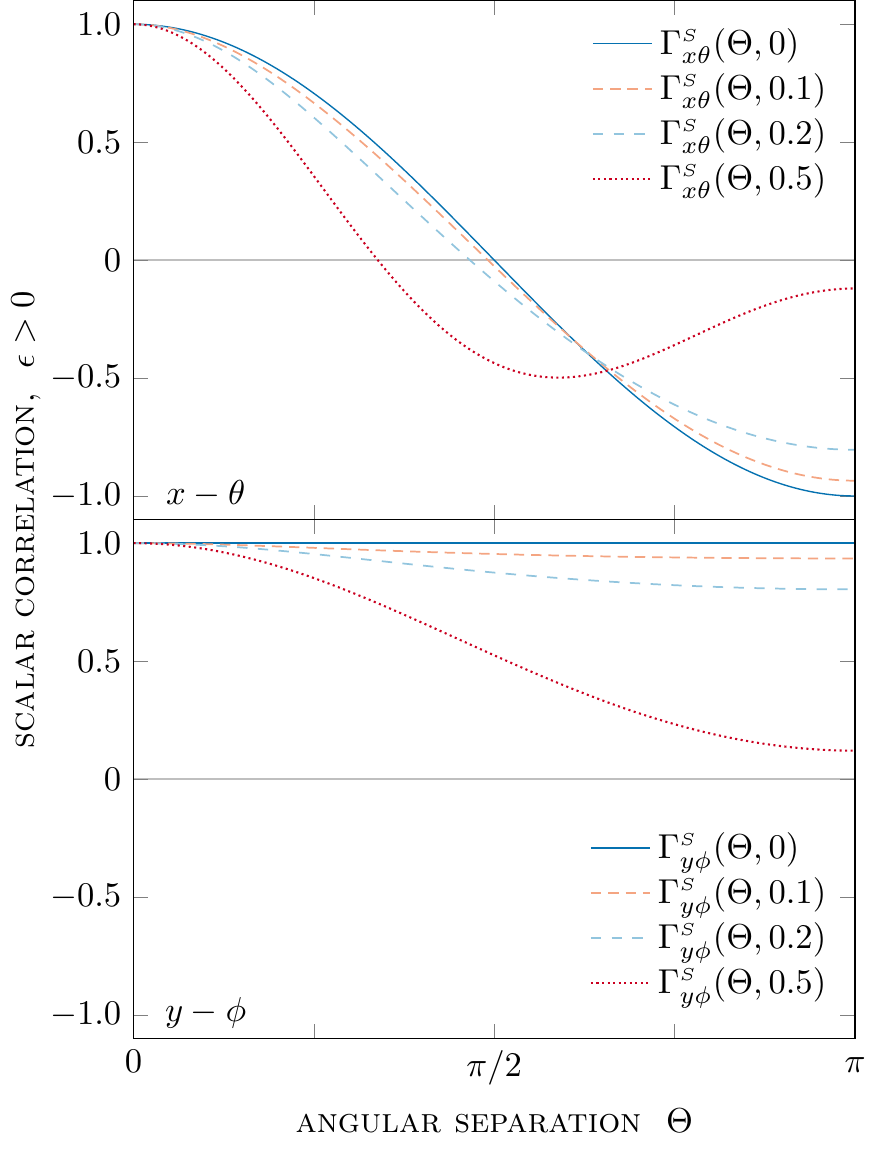}
\caption{The \(\G_{x\theta}^{\mathsc{S}}\) and \(\G_{y\phi}^{\mathsc{S}}\) cross-correlation functions for a background of sub-luminal (\(\e > 0\)) scalar transverse \textsc{gw}s for 4 different values of \(\e \in \{0, 0.1, 0.2, 0.5\}\) using the distant-source limit of the astrometric response. The \(y-\phi\) correlation in the \(\e = 0\) case is constant, which means the pattern is a pure dipole (\(\ell = 1\), see Fig.~\ref{fig:subluminalScalarCoeffs}); this symmetry breaks down as \(\e\) increases.}
\label{fig:subluminalScalarCorr}
\end{figure}

\begin{figure}[b]
\includegraphics[scale=1]{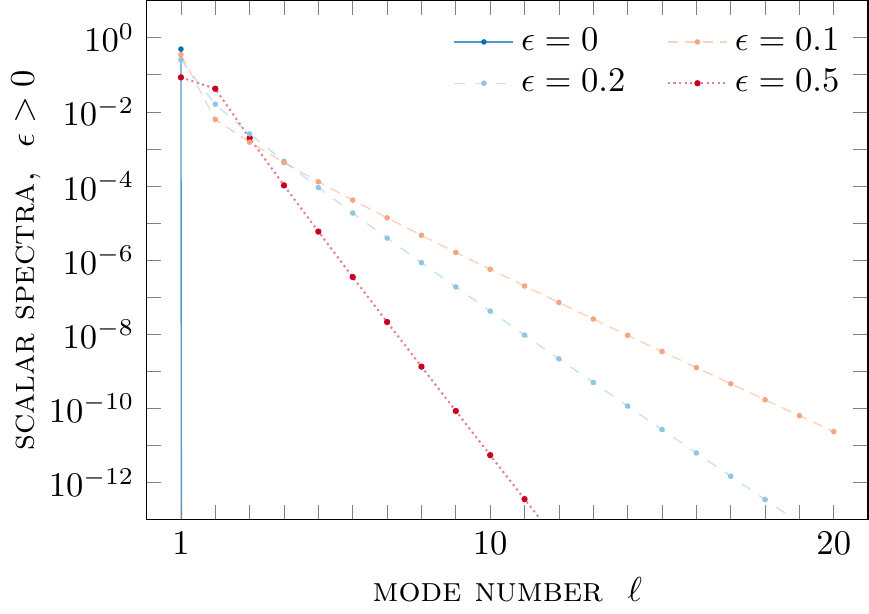}
\caption{The \(C_{\ell}^{\mathsc{S}}\) angular power spectra for a background of sub-luminal (\(\e > 0\)) transverse scalar \textsc{gw}s for 4 different values of \(\e~\!\!\in~\!\!\{0, 0.1, 0.2, 0.5\}\). The \(\e = 0\) spectrum has a single non-zero mode (\(\ell = 1\)), while non-zero \(\e\) excites higher-order multipoles, however increasing \(\e\) has the effect of concentrating the power at lower-order modes.}
\label{fig:subluminalScalarCoeffs}
\end{figure}

The current article uses the theoretical framework developed in \citep{2018PhRvD..97l4058M}, however alternative analytic approaches for obtaining these results exist. Notably, resolving the apparent proper motions of the stars along (spin-weighted) spherical harmonics has been used to address some of these problems before (especially in the context of \textsc{pta}s), and has been found to confirm the results presented here and in previous publications. The reader is invited to consult e.g. \citep{Qin:2018yhy} for introduction to the specifics of this method.

In this article the authors extend earlier calculations of \citep{2018PhRvD..97l4058M} and \cite{2018PhRvD..98b4020O} of the spatial correlation of astrometric deflections due to \textsc{gw} backgrounds of an arbitrary polarization to the case where the \textsc{gw}s propagate at non-luminal group velocities. These calculations are performed both in terms of the correlations between components of the astrometric deflection vector field on the sky and also in terms of the correlations between the coefficients in the vector spherical harmonic decomposition of the vector field; the relationship between these two equivalent descriptions is also clearly described. Section~\ref{sec:theory} introduces the theoretical background necessary to explain the results of the article. Subsection~\ref{sec:tensorial} discusses the results for the tensorial plus and cross polarizations (i.e. the familiar transverse traceless modes). The corresponding results for the alternative polarizations are presented in the following sections; the transverse scalar ``breathing mode'' is investigated in Subsection~\ref{sec:scalar}, the two vectorial modes are discussed in Subsection~\ref{sec:vectorial}, and the scalar longitudinal mode results are described in Subsection~\ref{sec:longitudinal}. Section~\ref{sec:additionalResults} presents the analytical results for redshift-redshift and redshift-astrometry correlations. The article concludes with a discussion of the results in Section~\ref{sec:discussion}.

\vspace*{-10pt}

\section{Theoretical background}
\label{sec:theory}
\vspace*{-10pt}
\noindent
A passing \textsc{gw} causes an \emph{astrometric deflection} -- a time-dependent shift in the apparent position of a star (or other distant light source). This effect was considered by many, including \citep{1996ApJ...465..566P, PhysRevD.83.024024, 2018PhRvD..97l4058M}; the notation of \citep{2018PhRvD..97l4058M} is used in this article. The astrometric deflection is a 2D vector dependent on the polarization, amplitude, direction and group velocity of the \textsc{gw}, as well as the direction and distance to the star; however, it will be most convenient to express it using 3D Cartesian coordinates (as a vector orthogonal to the direction to the star). The astrometric deflection is a linear function of the \textsc{gw} metric perturbation
\begin{align}\label{eq:astrometricShiftDef}
\delta n_{i} = \Re\{\tensor{\Delta}{_i^{jk}} h_{jk}\} ,
\end{align}
where \(h_{ij}\) is the \textsc{gw} metric perturbation at the Earth (\(\vect{x}\!=\!0\)), which is taken to be that of a plane wave, \(h_{ij} (t,\vect{x}) = \Re\{H_{ij}\exp(\rmi k_{\mu}x^{\mu})\}\). Hereafter, the restriction to the real part will be left implicit. Two expressions for the tensor \(\tensor{\Delta}{_i^{jk}}\) are presented here; each of them is applicable in a different regimes. The first, in eq.~(\ref{eq:astrometricShiftFull}), is the more general expression which depends on the direction of the unit vector \(\vect{n}\) and the distance \(d\) to the star (expressed in units of \textsc{gw} wavelengths), as well as on the direction to the source of the \textsc{gw}s \(\vect{q}\). For a more thorough derivation and discussion of this formula, including the result's applicability to cosmological space-times, the reader is encouraged to consult \citep{PhysRevD.83.024024} or \citep{2018PhRvD..97l4058M}; the latter of these follows the same formalism as the current article.
\begin{widetext}
\vspace*{-15pt}
\begin{align}
\label{eq:astrometricShiftFull}
\begin{split}
\tensor{\Delta}{_i^{jk}} (\vect{n}, \vect{q}, d) = & \;\,  \Bigg(\!\left\{1 + \dfrac{\rmi (2 - q^{r} n_{r})}{d (1 - q^{\ell} n_{\ell})}\,\Big(1 - \exp\!\big(-\rmi d (1 - q^{s} n_{s})\big)\Big)\!\right\} n_{i} \\[-5pt]
& \quad\quad\quad\quad \; - \left\{1 + \dfrac{\rmi}{d (1 - q^{\ell} n_{\ell})}\,\Big(1 - \exp\!\big(-\rmi d (1 - q^{s} n_{s})\big)\Big)\!\right\} q_{i}\Bigg) \dfrac{n^{j}n^{k}}{2(1 - q^{\ell} n_{\ell})} \\
& \quad\quad\quad\quad\quad\quad\quad\quad \vphantom{\left(\Bigg[\Bigg]^{2}\right)^{2}} \vphantom{\Bigg(\Bigg)^{2}} - \left\{\dfrac{1}{2} + \frac{\rmi}{d (1 - q^{\ell} n_{\ell})}\,\Big(1 - \exp\!\big(-\rmi d (1 - q^{s} n_{s})\big)\Big)\!\right\} n^{j} \delta_{i}^{k}
\end{split}
\end{align}
\vspace*{-10pt}
\end{widetext}

\begin{figure}[t]
\includegraphics[scale=1]{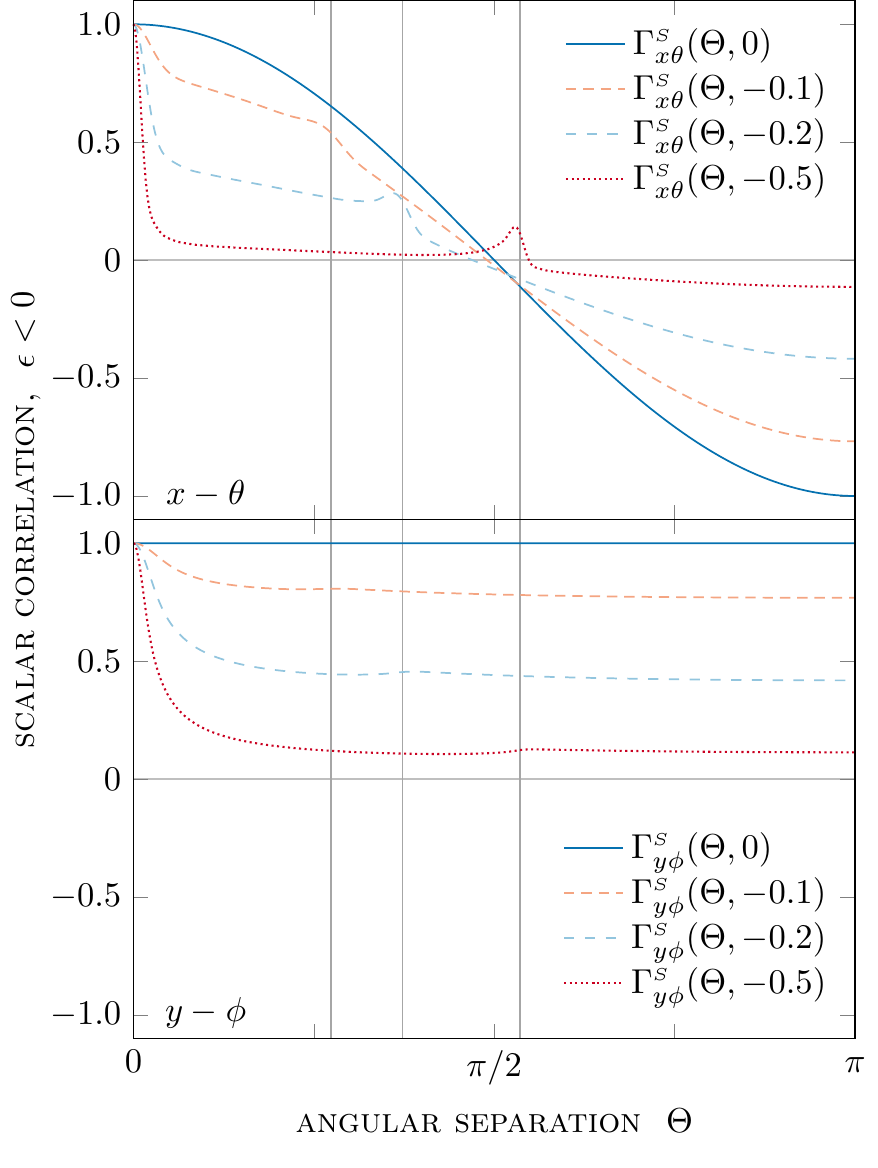}
\caption{The \(\G_{x\theta}^{\mathsc{S}}\) and \(\G_{y\phi}^{\mathsc{S}}\) cross-correlation functions for a background of super-luminal (\(\e < 0\)) scalar transverse \textsc{gw}s for 4 different values of \(\e~\in~\{0, -0.1, -0.2, -0.5\}\) using the full form of the astrometric response with \(d = 100\). The vertical lines mark the angles of the cones on which the astrometric response would be divergent if the distant-source limit formula was used. The \(y-\phi\) correlation in the \(\e = 0\) case is constant, which means the pattern is a pure dipole (\(\ell = 1\), see Fig.~\ref{fig:superluminalScalarCoeffs}); this symmetry breaks down as \(\e\) decreases.}
\label{fig:superluminalScalarCorr}
\end{figure}

\begin{figure}[t]
\includegraphics[scale=1]{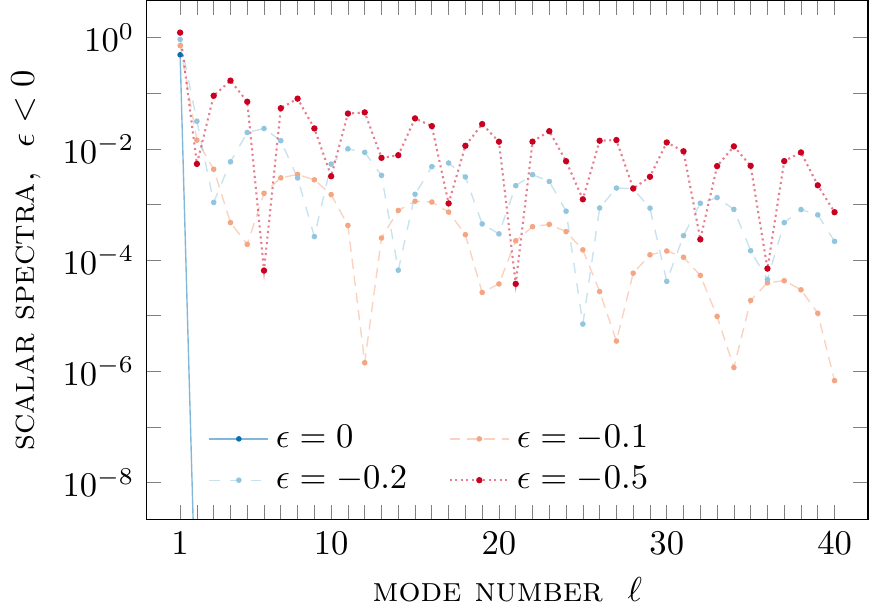}
\caption{The \(C_{\ell}^{\mathsc{S}}\) angular power spectra for a background of super-luminal (\(\e < 0\)) transverse scalar \textsc{gw}s for 4 different values of \(\e~\in~\{0, -0.1, -0.2, -0.5\}\). The \(\e = 0\) spectrum again has a single non-zero mode (\(\ell = 1\)), while non-zero \(\e\) excites higher-order multipoles, however decreasing \(\e\) has the effect of building up the power at higher-order modes. The oscillations which are visible in some of the spectra are due to the phases of the star terms in the astrometric shift formula.}
\label{fig:superluminalScalarCoeffs}
\end{figure}

The second expression for \(\tensor{\Delta}{_i^{jk}}\), in eq.~(\ref{eq:astrometricShiftLimit}), is the limiting value when the light source is a large distance away, if such a limit is well defined. In this case the expression simplifies and is independent of the distance parameter:
\begin{align}
\vspace*{-5pt}
\lim_{d \rightarrow \infty} \tensor{\Delta}{_i^{jk}} (\vect{n}, \vect{q}, d) = \frac{1}{2}\!\left(\!\frac{n_{i} - q_{i}}{1 - q^{\ell} n_{\ell}}\,n^{j} n^{k} - n^{j} \delta_{i}^{k}\right) .
\label{eq:astrometricShiftLimit}
\vspace*{-5pt}
\end{align}
Care must be taken when using the result for the distant-source limit in eq.~(\ref{eq:astrometricShiftLimit}); as discussed below this limit is not always well defined when the denominator \(1-\vect{q} \cdot \vect{n}\) vanishes. In this article, \(\vect{q}\) is opposite in direction to the \textsc{gw} wave 3-vector \(\vect{k}\) and is allowed to have a non-unit magnitude, signifying the \textsc{gw} travelling at non-luminal speed:
\begin{align}\label{eq:waveVectorRestriction}
k_{\mu} = - \omega \big(1, (1-\e)\,\vect{q}\big),
\end{align}
where the phase velocity of the \textsc{gw}s is \(v_{\rm{ph}}=1/(1-\e)\). In the case of massive gravity, with \(\e>0\), the group velocity is sub-luminal while the phase velocity is super-luminal. The two cases will be distinguished from here on by their group velocities: the case \(\e>0\) will be referred to as sub-luminal while \(\e<0\) will be referred to as super-luminal.

It is important to note that when \(\e > 0\) (sub-luminal \textsc{gw}s) the distant-source limit of the astrometric deflection given in eq.~(\ref{eq:astrometricShiftLimit}) is always well defined because \(1-\vect{q} \cdot \vect{n} > 0\). However, when \(\e < 0\) (super-luminal \textsc{gw}s) the distant-source limit is divergent when \(1-\vect{q} \cdot \vect{n} = 0\). Geometrically, this divergence takes the form of a ring on the sky, at an angle of \(\arcsec (1-\e)\) from the direction to the source of the \textsc{gw}s. This divergence is regularised by using instead the full astrometric response formula in eq.~(\ref{eq:astrometricShiftFull}), which includes the effects of a finite distance to the star. This is suitably illustrated in Appendix~\ref{sec:AstrometricShift}.

The astrometric response to a single monochromatic \textsc{gw} is given by eqs.~(\ref{eq:astrometricShiftFull}) and (\ref{eq:astrometricShiftLimit}). A stochastic background of \textsc{gw}s would produce a pattern of astrometric deflections which are highly correlated at large angular scales. This section summarises the formalism for deriving the correlation matrix for a background with an arbitrary polarization, and calculating the power spectra coefficients from them. The reader can find a more in-depth presentation of this in \citep{2018PhRvD..97l4058M}.

The \textsc{gw} perturbation due to a stochastic background of \textsc{gw}s may be decomposed and expressed as a sum of Fourier modes; the astrometric response to each Fourier mode can be expressed using eq.~\ref{eq:astrometricShiftDef} and the total response is given by
\begin{widetext}
\vspace*{-20pt}
\begin{align}
\delta n_{i} \left(\vect{n}, t\right) = \mathfrak{R} \left\{\sum_{\mathsc{P}}\!\int_{0}^{\infty} \!\!\!\!\!\!\dd f\,e^{-2\pi\rmi f t}\!\int_{S^{2}}\!\!\!\!\dd\Omega_{\vect{q}}\,A_{\mathsc{P}}\!\left(\vect{q}, f\right) \tensor{\Delta}{_{i}^{jk}}\!\left(\vect{n}, \vect{q}, d\right)\epsilon_{jk}^{\mathsc{P}}\!\left(\vect{q}\right)\!\right\}, \label{eq:CJMone}
\end{align}
\vspace*{-10pt}
\end{widetext}
where \(\tensor{\Delta}{_{i}^{jk}}\!\left(\vect{n}, \vect{q}, d\right)\) is given by eq.~(\ref{eq:astrometricShiftFull}), the sum over \(P\) includes all different \textsc{gw} polarizations, and the spatial integral is over the entire sphere of the sky. In general relativity, only the transverse \(+\) and \(\times\) modes are allowed, however modified theories of gravity could include up to 4 additional polarizations, the transverse scalar \(S\) mode, the vectorial \(X\) and \(Y\) modes, and the scalar longitudinal \(L\) mode. For a stochastic, Gaussian, zero-mean, stationary, isotropic, and unpolarized background the Fourier coefficients satisfy the following expectation relations:
\begin{subequations}
\begin{align}\label{eq:amplitudeCorrs}%
&\left<A_{\mathsc{P}}\!\left(\vect{q}, f\right)\right> = 0, \\
&\left<A_{\mathsc{P}}\!\left(\vect{q}, f\right) A^{*}_{\mathsc{P}'}\!\left(\vect{q}', f'\right) \right> = P(f)\,\delta_{\mathsc{PP'}}\,\delta_{S^{2}}(\vect{q},\vect{q}')\,\delta(f-f'),
\end{align}
\end{subequations}
where the angle brackets denote an average over all possible realisations of the \textsc{gw} background, and the function \(P(f)\) is related to the critical energy density of the Universe, \(\Omega_{\textsc{gw}} (f)\), via the normalization \citep{Allen:1997ad, 1993PhRvD..48.2389F}
\begin{align}
P(f) = \frac{3 H_{0}^{2}\,\Omega_{\textsc{gw}} (f)}{8 \pi^{3} f^{3}}.
\end{align}

Consider two stars at positions on the sky \(\vect{n}\) and \(\vect{m}\). The astrometric deflection of each star is a linear function of the \textsc{gw} perturbation (cf. eq.~(\ref{eq:astrometricShiftDef})) and therefore is also a zero-mean Gaussian random variable whose statistics are fully specified by the two-point expectation. The two-point expectation of the astrometric deflection factorises into a term that depends on the relative times of the measurements \(T(t,t')\) and a geometric factor which depends on the locations of the stars on the sky \(\Gamma_{ij} (\vect{n}, \vect{m})\);
\begin{align}\label{eq:new1}
\left<\delta n_{i} (\vect{n}, t)\,\delta n_{j} (\vect{m}, t')\right> = T(t,t')\,\sum_{\mathsc{P}}\Gamma^{\mathsc{P}}_{ij} (\vect{n}, \vect{m})\,.
\end{align}
\(T\) is defined explicitly in eq.~(31) of \citep{2018PhRvD..97l4058M} but is not needed here. It is only the geometric factor \(\Gamma\) that is interesting here as it is only this part that depends on the details of the \textsc{gw} polarization and propagation speed.

While eq.~(\ref{eq:new1}) is conceptually the necessary result, in practice it is useful to find a geometrically intuitive interpretation of this correlation matrix. Since the astrometric deflection lies in the tangent space of the sphere, it can be fully specified by 2 orthogonal components, and therefore one of the dimensions of the \(3 \times 3\) matrix on the right-hand side of eq.~(\ref{eq:new1}) is redundant. The statistical rotational invariance of the background can be used to set \(\vect{n} = (0, 0, 1)\) and \(\vect{m} = (\sin(\Theta), 0, \cos(\Theta))\), with \(\Theta = \arccos (\vect{n} \cdot \vect{m})\). If the astrometric response at each of the two stars is resolved parallelly and perpendicularly to the arc connecting the stars on the sphere, the correlation becomes a \(2 \times 2\) matrix instead (cf. Fig.~(4) of \citep{2018PhRvD..97l4058M}). In this gauge, in the case of an isotropic background the off-diagonal entries of the matrix vanish, and the remaining correlation functions depend solely on the angular separation of the two stars \(\Theta\) (when \(\tensor{\Delta}{_{i}^{jk}}\) is given by eq.~(\ref{eq:astrometricShiftLimit})). Thus, \(\Gamma^{\mathsc{P}}_{ij} (\vect{n}, \vect{m})\) reduces to
\begin{align}\label{eq:corrMatrix}
\G^{\mathsc{P}} (\vect{n}, \vect{m}) = \left(\begin{matrix}
\G_{x\theta}^{\mathsc{P}} (\Theta) & 0 \\
0  & \G_{y\phi}^{\mathsc{P}} (\Theta) \\
\end{matrix}\right).
\end{align}
The functions \(\G_{ab} (\Theta)\) are obtained by integrating components of the astrometric deflection over the sphere of all possible directions to the \textsc{gw} source:
\begin{subequations}\label{eq:GammaDef}
\begin{align}
\Gamma^{\mathsc{P}}_{x\theta}(\Theta) = \int_{S^{2}} \!\!\!\dd\Omega_{\vect{q}}\;\delta n_{x}^{\mathsc{P}}(\vect{q})\,\delta m_{\theta}^{\mathsc{P}}(\vect{q}) \,, \label{eq:new2}\\ \nonumber\\[-10pt]
\Gamma^{\mathsc{P}}_{y\phi}(\Theta) = \int_{S^{2}} \!\!\!\dd\Omega_{\vect{q}}\;\delta n_{y}^{\mathsc{P}}(\vect{q})\,\delta m_{\phi}^{\mathsc{P}}(\vect{q}) \label{eq:new3}\,,
\end{align}
\end{subequations}
where the shorthands \(\delta n_{a} (\vect{n}, \vect{q}) \equiv \delta n_{a} (\vect{q})\) and \(\delta n_{b} (\vect{m}, \vect{q}) \equiv \delta m_{b} (\vect{q})\) have been used. The components \(\delta n^{\mathsc{P}}_{x}\), \(\delta n^{\mathsc{P}}_{y}\), \(\delta m^{\mathsc{P}}_{\theta}\), and \(\delta m^{\mathsc{P}}_{\phi}\) are defined in eqs.~(41) and (42) of \cite{2018PhRvD..97l4058M}. The two correlation functions \(\Gamma^{\mathsc{P}}_{x\theta}\) and \(\Gamma^{\mathsc{P}}_{y\phi}\) have the geometric significance of describing the correlation between the components of the astrometric deflection vectors which lie parallel and perpendicular respectively to the great circle joining \(\vect{n}\) and \(\vect{m}\). Alternatively, these two functions can be thought of as describing the divergence and curl parts of the astrometric deflection vector field respectively. 
For a detailed description of the geometric construction the reader is referred to \citep{2018PhRvD..97l4058M}.

In this article, the functions from eq.~(\ref{eq:GammaDef}) are derived, plotted, and examined in detail for positive and negative values of \(\e\): \(\G_{ab}^{+, \times}\) in Figs.~\ref{fig:subluminalTensorialCorr} and \ref{fig:superluminalTensorialCorr}, \(\G_{ab}^{\mathsc{S}}\) in Figs.~\ref{fig:subluminalScalarCorr} and \ref{fig:superluminalScalarCorr}, \(\G_{ab}^{\mathsc{X}, \mathsc{Y}}\) in Figs.~\ref{fig:subluminalVectorialCorr} and \ref{fig:superluminalVectorialCorr}, and \(\G_{ab}^{\mathsc{L}}\) in Figs.~\ref{fig:subluminalLongitudinalCorr} and \ref{fig:superluminalLongitudinalCorr}, respectively.

Instead of dealing with the astrometric deflection vector field directly, it may be convenient to decompose it into \textsc{vsh}. This has been considered for regular \textsc{gw}s in general relativity by \cite{PhysRevD.83.024024} and for alternative \textsc{gw} polarizations (travelling at the speed of light) by \cite{2018PhRvD..98b4020O}. Any vector field on the surface of a sphere can be decomposed into a sum of ``electric'' \(E\), and ``magnetic'' \(B\) \textsc{vsh} functions (where the conventions in e.g. \citep{garfken67:math} are used)
\begin{align}
\delta n_{i}^{\mathsc{P}}(\vect{n}, \vect{q}) = \sum_{\mathsc{Q\!\in\!\{\!E,\!B\!\}}} \sum_{\ell = 1}^{\infty} \sum_{m = -\ell}^{\ell} a_{\ell m}^{\mathsc{PQ}} (\vect{q})\;Y_{\ell mi}^{\mathsc{Q}} (\vect{n}).
\end{align}
Because the transformation between the \(\delta n_{i}^{\mathsc{P}}(\vect{n},\vect{q})\) and expansion coefficients \(a^{\mathsc{PQ}}_{lm}(\vect{q})\) is linear, the coefficients are also distributed as a zero-mean Gaussian random variable. The two-point expectation of the expansion coefficients is given by (as quoted by e.g. \cite{2018PhRvD..98b4020O}), 
\begin{align} \label{eq:newLABEL1}
\left<a_{lm}^{\mathsc{P}\mathsc{Q}}\,a_{l'm'}^{\mathsc{P'}\mathsc{Q'}^{*}}\right> = C_{\ell}^{\mathsc{P,} \mathsc{Q}}\,\delta_{\mathsc{PP'}}\,\delta_{\mathsc{QQ'}}\,\delta_{ll'}\,\delta_{mm'}.
\end{align}
The coefficients \(C_{\ell}^{\mathsc{P,} \mathsc{Q}}\) are given by (valid individually for each polarization \(P\))
\begin{widetext}
\vspace*{-20pt}
\begin{subequations}\label{eq:spectraCoeffsFormulae}
\begin{align}
C_{\ell}^{\mathsc{E}} = \frac{1}{2\ell (\ell + 1)} \bigintsss_{0}^{\pi} \!\!\!\!\dd\Theta \, \sin\Theta \, P_{\ell}\!\left(\cos\Theta\right) \left(\!-\G_{x\theta}^{\prime\prime}(\Theta) + \frac{1}{\sin\Theta}\,\G_{y\phi}^{\prime} (\Theta) - 2\,\frac{\cos\Theta}{\sin\Theta}\,\G_{x\theta}^{\prime} (\Theta) + \G_{x\theta} (\Theta)\!\right) \phantom{,} \\[5pt]
C_{\ell}^{\mathsc{B}} = \frac{1}{2\ell (\ell + 1)} \bigintsss_{0}^{\pi} \!\!\!\!\dd\Theta \, \sin\Theta \, P_{\ell}\!\left(\cos\Theta\right) \left(\!-\G_{y\phi}^{\prime\prime}(\Theta) + \frac{1}{\sin\Theta}\,\G_{x\theta}^{\prime} (\Theta) - 2\,\frac{\cos\Theta}{\sin\Theta}\, \G_{y\phi}^{\prime} (\Theta) + \G_{y\phi} (\Theta)\!\right),
\end{align}
\end{subequations}
\end{widetext}
where the primes denote derivatives with respect to \(\Theta\). They measure the power stored in a particular \textsc{vsh} \(Y^{\mathsc{Q}}_{lm}\), while the isotropy of the \textsc{gw} background ensures there is no dependence on the \(m\) index. These expressions are equivalent to the ones given in \citep{2018PhRvD..98b4020O}, but include a corrected normalization factor.

It is important to explain that eqs.~(\ref{eq:GammaDef}) and (\ref{eq:spectraCoeffsFormulae}) are equivalent representations of the same information, and are thus interchangeable. The former describes the correlation pattern of astrometric deflections by specifying the correlations between their components as a function of the angular separation on the sky, \(\Theta\). The latter describes the same correlated pattern by computing the correlations between the coefficients of the \textsc{vsh} decomposition of the vector field of astrometric deflections on the sky.

In practice, \(C_{\ell}^{\mathsc{E}}\) and \(C_{\ell}^{\mathsc{B}}\) are combined to produce a single spectrum which quantifies the angular power distribution of a correlation function,
\begin{align}\label{eq:coeffsDef}
C_{\ell} = \sqrt{\frac{(C_{\ell}^{\mathsc{E}})^{2} + (C_{\ell}^{\mathsc{B}})^{2}}{2}},
\end{align}
which is valid separately for each polarization content \(P\). In this article, the coefficients from eq.~(\ref{eq:coeffsDef}) are plotted and examined in detail for positive and negative values of \(\e\): \(C_{\ell}^{+, \times}\) in Figs.~\ref{fig:subluminalTensorialCoeffs} and \ref{fig:superluminalTensorialCoeffs}, \(C_{\ell}^{\mathsc{S}}\) in Figs.~\ref{fig:subluminalScalarCoeffs} and \ref{fig:superluminalScalarCoeffs}, \(C_{\ell}^{\mathsc{X}, \mathsc{Y}}\) in Figs.~\ref{fig:subluminalVectorialCoeffs} and \ref{fig:superluminalVectorialCoeffs}, and \(C_{\ell}^{\mathsc{L}}\) in Figs.~\ref{fig:subluminalLongitudinalCoeffs} and \ref{fig:superluminalLongitudinalCoeffs}, respectively.

The statistics of the stochastic pattern of astrometric deflections are completely described by either eqs.~(\ref{eq:new1}) or eq.~(\ref{eq:newLABEL1}). The former describes the correlations in terms of the components of the astrometric deflection vector field while the later describes the correlations in terms of the coefficients of the \textsc{vsh} expansion. The two descriptions are equivalent and eqs.~(\ref{eq:spectraCoeffsFormulae}) describe the relationships between the two.

\section{Non-luminal astrometric\texorpdfstring{\\}{}correlation functions and\texorpdfstring{\\}{}power spectra}
\label{sec:non-luminalCorrelations}
\vspace*{-10pt}
\noindent
This section examines the sub- and super-luminal cases of the four relevant polarization states of a \textsc{gw} background: tensorial, scalar transverse, vectorial, and scalar longitudinal. For each of these four cases, the correlations functions for \(\e>0\) and \(\e<0\) are presented, compared, and discussed. The corresponding angular power spectra are also shown, and the effects of the value of \(\e\) are explained.

\vspace*{-15pt}

\subsection{Tensorial Transverse-Traceless Polarizations}
\label{sec:tensorial}
\vspace*{-15pt}
\noindent
The most familiar results are those for transverse-traceless \textsc{gw} backgrounds, for which \(P \in \{+, \times\}\). These are the backgrounds expected to be found in \textsc{gr}, and as such it is interesting to examine them in the context of probing the speed of gravity. The \(\e = 0\) case was investigated in detail in \citep{2018PhRvD..97l4058M}, here it is used as a reference point to establish the effect of positive and negative values of \(\e\).

For \(\e > 0\), Appendix~\ref{sec:evalTensorialFunctions} explains how to evaluate the integrals \(\G_{x\theta}^{+} (\Theta, \e)\), \(\G_{y\phi}^{+} (\Theta, \e)\), \(\G_{x\theta}^{\times} (\Theta, \e)\), and \(\G_{y\phi}^{\times} (\Theta, \e)\) defined by eqs.~(\ref{eq:GammaDef}). The spatial correlation matrix in a background with multiple polarizations is the sum of individual spatial correlations, hence \(\G_{x\theta}^{+, \times} (\Theta, \e) = \G_{x\theta}^{+} (\Theta, \e) + \G_{x\theta}^{\times} (\Theta, \e)\), and similarly for \(\G_{y\phi}^{+, \times} (\Theta, \e)\). The two functions \(\G_{x\theta}^{+, \times} (\Theta, \e)\) and \(\G_{y\phi}^{+, \times} (\Theta, \e)\) are given explicitly below in eqs.~(\ref{eq:subluminalTensorialxthCorr}) and (\ref{eq:subluminalTensorialyphCorr}). All correlation function results in this article are expressed in terms of \(\s = \sin(\Theta/2)\).
\vspace*{-5pt}
\begin{widetext}
\begin{subequations}
\begin{align}\label{eq:subluminalTensorialxthCorr}
\begin{split}
&\G_{x\theta}^{+, \times} (\Theta, \e) = \frac{2\pi}{3}\,\frac{\left(1-6\e-\e^{2}+4\e^{3}-\e^{4}\right) - \left(7-20\e+10\e^{2}\right)\!\s[2]}{(1-\e)^{4}} \\
&\RepQuad{5}+\frac{\pi}{2}\,\frac{2\e^{2}(2-\e)^{2} + \e(2-\e)\!\left(4-14\e+7\e^{2}\right)\!\s[2] + 2\!\left(4-12\e+18\e^{2}-12\e^{3}+3\e^{4}\right)\!\s[4]}{(1-\e)^{5}\!\left(1-\s[2]\right)}\,\eln \\
&\RepQuad{5}-\pi\,\frac{\left(\e^{2}(2-\e)^{2} + 8\e(2-\e)(1-\e)^{2}\s[2] + 8(1-\e)^{4}\s[4]\right)\!\s}{(1-\e)^{5}\!\left(1-\s[2]\right)\!\ssqrt}\times \\
&\RepQuad{28} \times\sln
\end{split}\\[10pt]\label{eq:subluminalTensorialyphCorr}
\begin{split}
&\G_{y\phi}^{+, \times} (\Theta, \e) = \frac{\pi}{3}\,\frac{\left(2-12\e+10\e^{2}-4\e^{3}+\e^{4}\right) - 2(1-\e)^{2}\!\left(7-2\e+\e^{2}\right)\!\s[2]}{(1-\e)^{4}}\\
&\RepQuad{5}+ \frac{\pi}{2}\,\frac{2\e^{2}(2-\e)^{2} + \e(2-\e)\!\left(12-26\e+13\e^{2}\right)\!\s[2] + 4(1-\e)^{2}\!\left(2-6\e+3\e^{2}\right)\!\s[4]}{(1-\e)^{5}\!\left(1-\s[2]\right)}\,\eln \\
&\RepQuad{5}- \pi\,\frac{\left(\e^{2}(2-\e)^{2} + 8\e(2-\e)(1-\e)^{2}\s[2] + 8(1-\e)^{4}\s[4]\right)\!\ssqrt}{(1-\e)^{5}\!\left(1-\s[2]\right)}\times \\
&\RepQuad{28} \times\sln \\[10pt]
\end{split}
\end{align}
\end{subequations}
\end{widetext}

\begin{figure}[b]
\includegraphics[scale=1]{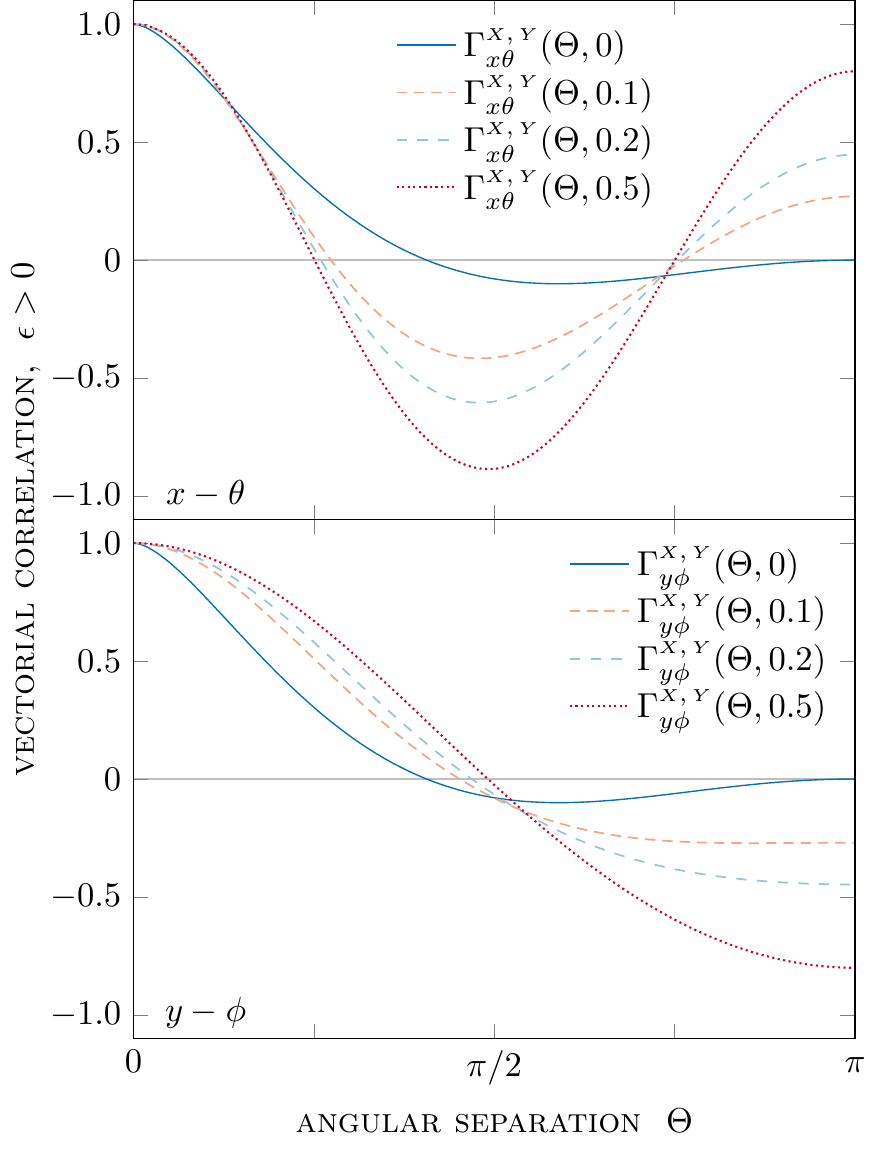}
\caption{The \(\G_{x\theta}^{\mathsc{X,} \mathsc{Y}}\) and \(\G_{y\phi}^{\mathsc{X,} \mathsc{Y}}\) cross-correlation functions for a background of sub-luminal (\(\e > 0\)) vectorial \textsc{gw}s for 4 different values of \(\e \in \{0, 0.1, 0.2, 0.5\}\) using the distant-source limit of the astrometric response. While the \(x-\theta\) and \(y-\phi\) curves are the same in the \(\e = 0\) case, the degeneracy breaks down for any \(\e > 0\). The angular power spectra are shown in Fig.~\ref{fig:subluminalVectorialCoeffs}.}
\label{fig:subluminalVectorialCorr}
\end{figure}

\begin{figure}[b]
\includegraphics[scale=1]{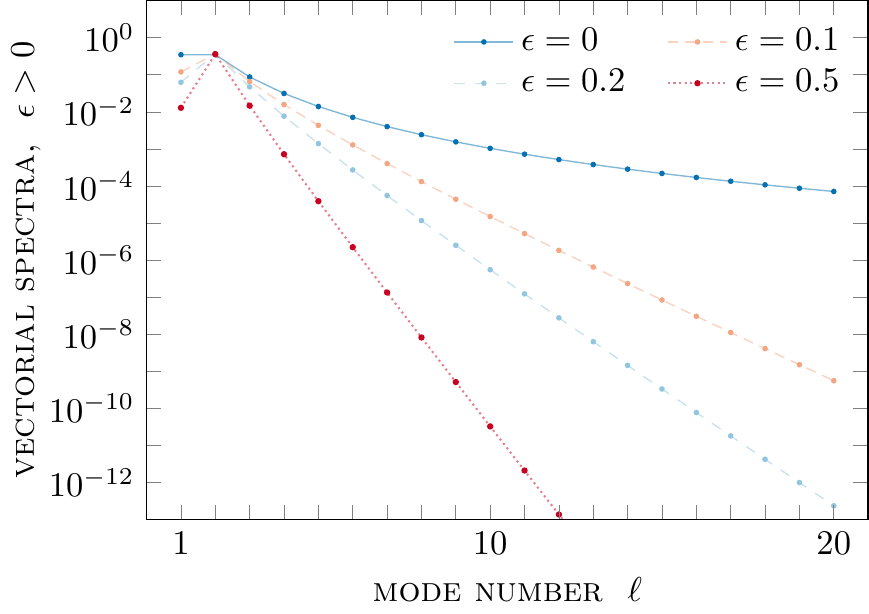}
\caption{The \(C_{\ell}^{\mathsc{X,} \mathsc{Y}}\) angular power spectra for a background of sub-luminal (\(\e>0\)) vectorial \textsc{gw}s for 4 different values of \(\e \in \{0, 0.1, 0.2, 0.5\}\). The \(\e = 0\) case is equal parts dipole and quadrupole, while increasing \(\e\) has the effect of suppressing the power at higher multipoles (and of the \(\ell = 1\) mode), making the spectra more quadrupolar.}
\label{fig:subluminalVectorialCoeffs}
\end{figure}

\begin{figure}[t]
\includegraphics[scale=1]{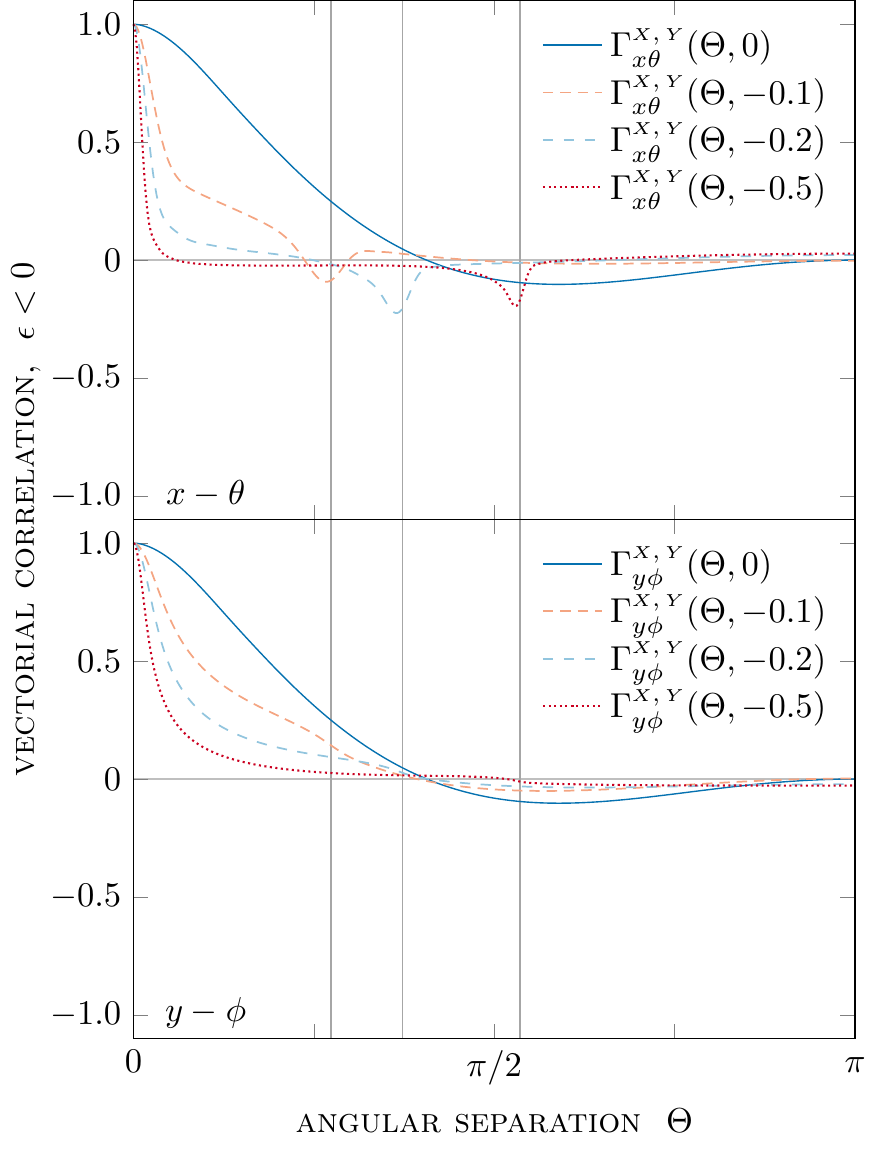}
\caption{The \(\G_{x\theta}^{\mathsc{X,} \mathsc{Y}}\) and \(\G_{y\phi}^{\mathsc{X,} \mathsc{Y}}\) cross-correlation functions for a background of super-luminal (\(\e < 0\)) vectorial \textsc{gw}s for 4 different values of \(\e~\in~\{0, -0.1, -0.2, -0.5\}\) using the full form of the astrometric response with \(d = 100\). The vertical lines mark the angles of the cones on which the astrometric response would be divergent if the distant-source limit formula was used. See Fig.~\ref{fig:superluminalVectorialCoeffs} for the angular power spectra of these functions.}
\label{fig:superluminalVectorialCorr}
\end{figure}

\begin{figure}[t]
\includegraphics[scale=1]{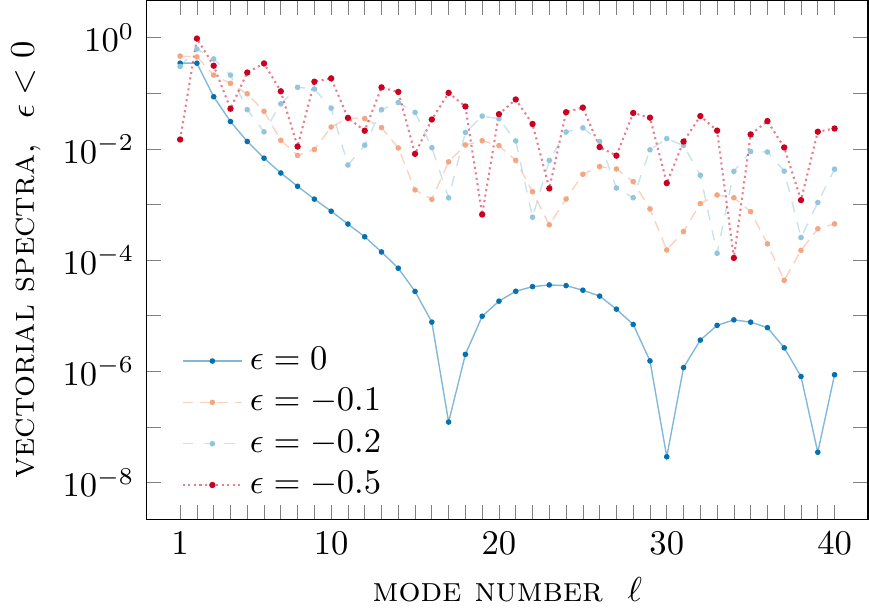}
\caption{The \(C_{\ell}^{\mathsc{X,} \mathsc{Y}}\) angular power spectra for a background of super-luminal (\(\e<0\)) vectorial \textsc{gw}s for 4 different values of \(\e \in \{0, -0.1, -0.2, -0.5\}\). A lower value of \(\e\) has the effect of increasing the power at higher multipoles. The oscillations which are visible in all of the spectra are due to the phases of the star terms in the astrometric shift formula.}
\label{fig:superluminalVectorialCoeffs}
\end{figure}

Whilst these two functions were identical for luminal \textsc{gw} backgrounds (cf. eq.~(45) of \citep{2018PhRvD..97l4058M}), one effect of a non-zero \(\e\) is that the \(x-\theta\) and \(y-\phi\) correlations are now not equal to each other. This is illustrated in a plot in Fig.~\ref{fig:subluminalTensorialCorr}, for four different values of \(\e\): \(0, 0.1, 0.2,\) and \(0.5\). The overall shape of the curves evolve as the value of \(\e\) is increased from 0. This is the effect which could in principle be utilized in order to place a constraint on the speed of gravity using measurements of the astrometric patterns due to a stochastic gravitational wave background.

\begin{figure}[b]
\includegraphics[scale=1]{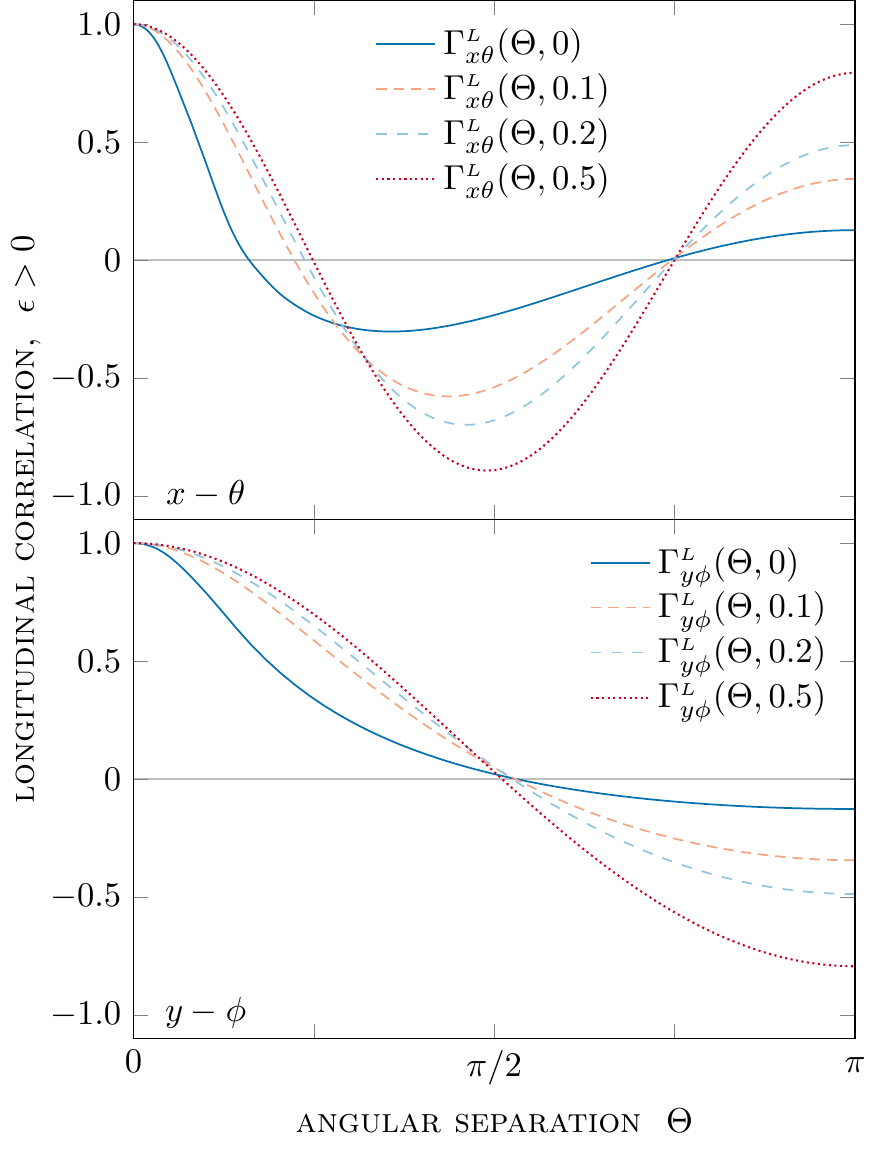}
\caption{The \(\G_{x\theta}^{\mathsc{L}}\) and \(\G_{y\phi}^{\mathsc{L}}\) cross-correlation functions for a background of sub-luminal (\(\e > 0\)) scalar longitudinal \textsc{gw}s for 4 different values of \(\e \in \{0, 0.1, 0.2, 0.5\}\) using the distant-source limit of the astrometric response. The angular power spectra of these functions are shown in Fig.~\ref{fig:subluminalLongitudinalCoeffs}.}
\label{fig:subluminalLongitudinalCorr}
\end{figure}

\begin{figure}[b]
\includegraphics[scale=1]{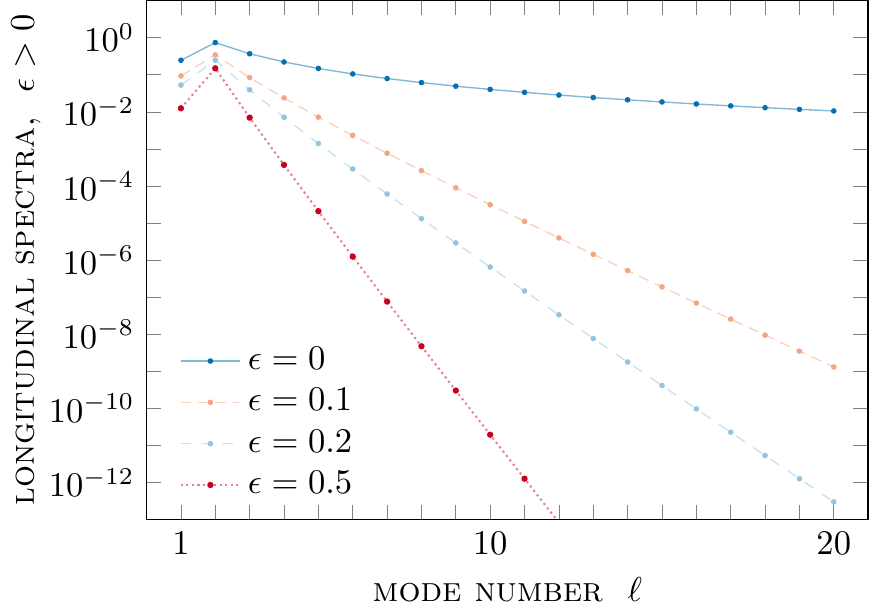}
\caption{The \(C_{\ell}^{\mathsc{L}}\) angular power spectra for a background of sub-luminal (\(\e>0\)) scalar longitudinal \textsc{gw}s for 4 different values of \(\e \in \{0, 0.1, 0.2, 0.5\}\). Increasing \(\e\) has the effect of suppressing the power at higher multipoles (and of the \(\ell = 1\) mode), making the spectra more quadrupolar.}
\label{fig:subluminalLongitudinalCoeffs}
\end{figure}

As an alternative way to quantify the differences between those correlation functions, the angular power spectra \(C_{\ell}\) for eqs.~(\ref{eq:subluminalTensorialxthCorr}) and (\ref{eq:subluminalTensorialyphCorr}) can be computed through eqs.~(\ref{eq:spectraCoeffsFormulae}). In the case \(\e = 0\), a closed form expression can be derived, and the result is given in Appendix~\ref{app:tensorialLuminalSpectra}. For \(\e > 0\), even though analytical expressions for the coefficients can be found, it is computationally cheaper to calculate them numerically. The values obtained are shown in Fig.~\ref{fig:subluminalLongitudinalCoeffs}.

\begin{figure}[t]
\includegraphics[scale=1]{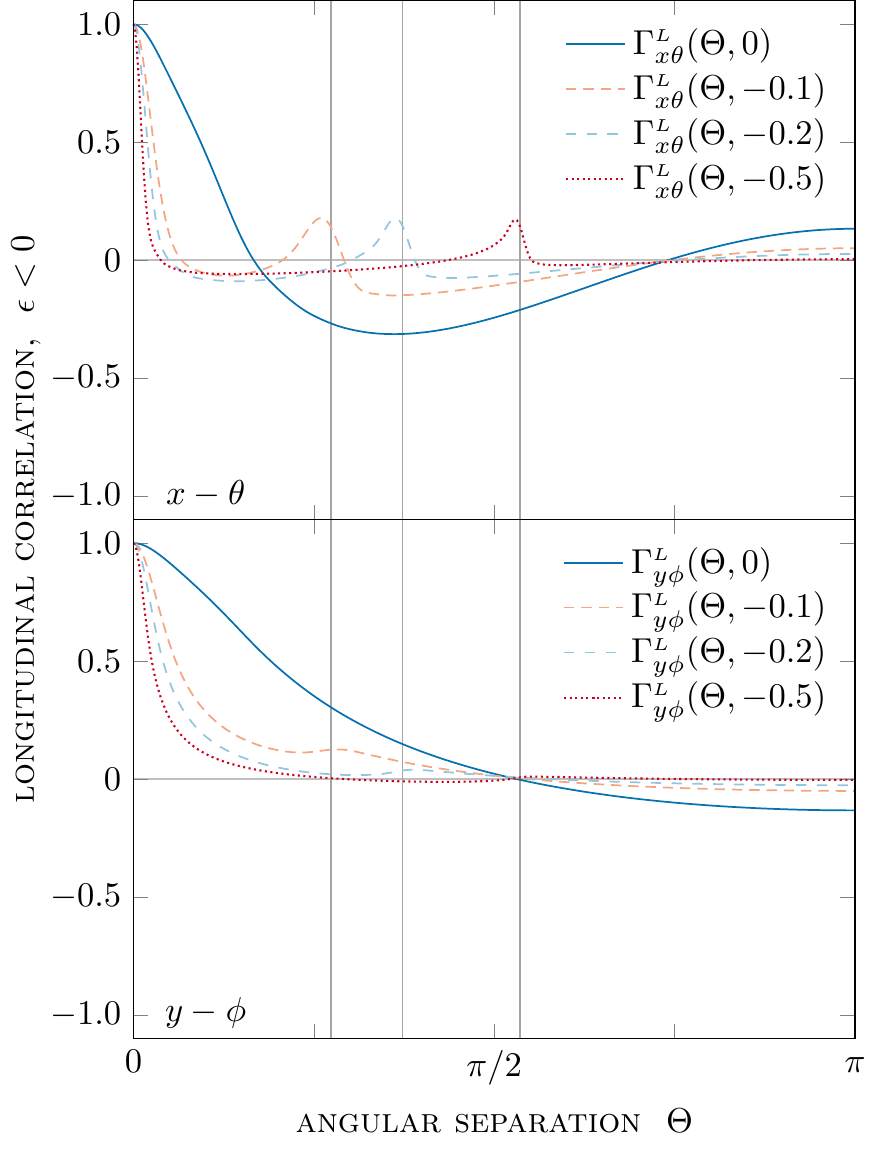}
\caption{The \(\G_{x\theta}^{\mathsc{L}}\) and \(\G_{y\phi}^{\mathsc{L}}\) cross-correlation functions for a background of super-luminal (\(\e<0\)) longitudinal scalar \textsc{gw}s for 4 different values of \(\e \in \{0, -0.1, -0.2, -0.5\}\) using the full form of the astrometric response with \(d = 100\). The vertical lines mark the angles of the cones on which the astrometric response would be divergent if the distant-source limit formula was used. See Fig.~\ref{fig:superluminalLongitudinalCoeffs} for the angular power spectra of these functions.}
\label{fig:superluminalLongitudinalCorr}
\end{figure}

\begin{figure}[t]
\includegraphics[scale=1]{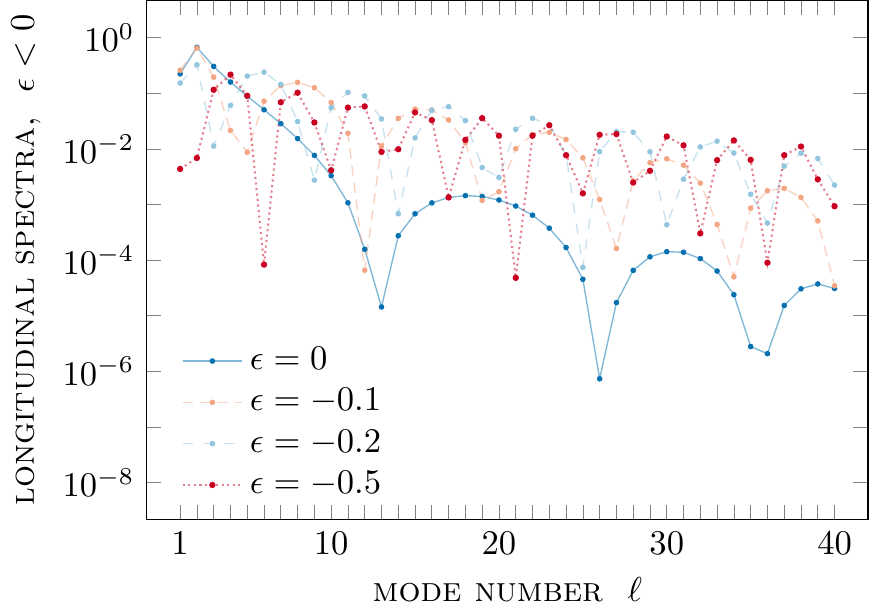}
\caption{The \(C_{\ell}^{\mathsc{L}}\) angular power spectra for a background of super-luminal (\(\e<0\)) longitudinal scalar \textsc{gw}s for 4 different values of \(\e \in \{0, -0.1, -0.2, -0.5\}\). A lower value of \(\e\) has the effect of increasing the power at higher multipoles. The oscillations which are visible in all of the spectra are due to the phases of the star terms in the astrometric shift formula.}
\label{fig:superluminalLongitudinalCoeffs}
\end{figure}

In the case of \(\e < 0\), as explained before, the correlation function integral eqs.~(\ref{eq:GammaDef}) can no longer be evaluated with the distant-source limit of the astrometric deflection. Instead, one has to use the exact formula which includes the ``star term'' perturbation by invoking the distance to the star \(d\). One consequence is that the integrals in eqs.~(\ref{eq:GammaDef}) cannot be evaluated analytically any longer, and one has to resort to numerical tools in order to obtain the results in this case. Some details  of the calculation are offered in Appendix~\ref{sec:evalTensorialFunctions}, and the results are shown in Fig.~\ref{fig:superluminalTensorialCorr} for the mirrored 4 values of \(\e\): \(0, -0.1, -0.2,\) and \(-0.5\). The difference between the \(\e = 0\) curve in this case and in Fig.~\ref{fig:subluminalTensorialCorr} is that here the ``star'' term is included in the astrometric responses for both stars, with \(d = 100\) wavelengths.

The angular power spectrum coefficients in the case \(\e < 0\) cannot be calculated analytically either, but can be computed numerically through eqs.~(\ref{eq:spectraCoeffsFormulae}). The four power spectra corresponding to the curves in Fig.~\ref{fig:superluminalTensorialCorr} were calculated numerically and are presented in Fig.~\ref{fig:superluminalTensorialCoeffs}.

Special attention ought to be paid to the relationship between the results in Figs.~\ref{fig:subluminalTensorialCorr} and \ref{fig:subluminalTensorialCoeffs} (and Figs.~\ref{fig:superluminalTensorialCorr} and \ref{fig:superluminalTensorialCoeffs} alike). The correlation curve plots fully describe the correlated pattern of astrometric deflections by giving the correlation between particular components of the astrometric deflection vector field as a function \(\Gamma_{ab}^{\mathsc{P}}\) of the angular separation on the sky, \(\Theta\). The angular power spectra plots describe the correlated pattern of astrometric deflections by specifying the correlations between the coefficients in the \textsc{vsh} decomposition of the astrometric deflection vector field. The two descriptions are of course completely equivalent and interchangeable, and eqs.~(\ref{eq:spectraCoeffsFormulae}) give the necessary relations for converting from the former to the latter.

It is expected that \textsc{gr} is the correct description of gravity, and therefore the measured correlation patterns should correspond to the ones predicted here. In particular, the astrometric deflection correlation curve should be the blue curve plotted in Fig.~\ref{fig:subluminalTensorialCorr}. However, if the speed of gravity does not match the speed of light (\(\e \neq 0\)), there would be a discrepancy between the actual correlation curve and the blue curve (\(\e = 0\)) in Fig.~\ref{fig:subluminalTensorialCorr}: the new correlation would look like one of the other curves in Figs.~\ref{fig:subluminalTensorialCorr} or \ref{fig:superluminalTensorialCorr}. This change could equivalently be analysed from the spectral point of view. The correlation spectrum would change: the power would shift towards (away from) \(\ell = 2\) for sub-(super-)luminal \textsc{gw}s; see the plots in Figs.~\ref{fig:subluminalTensorialCoeffs} and \ref{fig:superluminalTensorialCoeffs}.

\vspace*{-5pt}
\subsection{Scalar Transverse (Breathing)\texorpdfstring{\\}{}Polarization}
\label{sec:scalar}
\noindent
The most prominent candidates for a modified theory of gravity are scalar-tensor theories, for which the force of gravity is mediated by a combination of the metric (like in \textsc{gr}) plus a real scalar field. Such theories usually have an additional \textsc{gw} polarization degree of freedom, in addition to the two transverse-traceless states of \textsc{gr}. For the following discussion, it is assumed that the \textsc{gw} background consists solely of transverse scalar \(S\) mode \textsc{gw}s.

The analysis in this case is similar to the one in the previous section. The correlation functions for \(\e > 0\) are given by eqs.~(\ref{eq:SublimunalScalarxthCorr}) and (\ref{eq:SublimunalScalaryphCorr}) in Appendix~\ref{sec:evalScalarFunctions} along with brief notes on their derivation, and are plotted in Fig.~\ref{fig:subluminalScalarCorr}. It is important to note that the \(y-\phi\) correlation function, which used to be constant for luminal isotropic backgrounds (cf. eq.~(47b) from \citep{2018PhRvD..97l4058M}), is now variable with the angular separation \(\Theta\), corresponding to breaking the purely dipole pattern. The angular power spectra are shown in Fig.~\ref{fig:subluminalScalarCoeffs}, with the luminal case considered in more detail in Appendix~\ref{app:scalarLuminalSpectra}. As expected, for \(\e = 0\) only the \(\ell = 1\) coefficient is non-zero (corresponding to a dipole pattern), while for \(\e > 0\) power is re-distributed to higher-order multipoles.

For \(\e<0\), the correlation functions cannot be computed analytically, only numerically; they are plotted in Fig.~\ref{fig:superluminalScalarCorr} and a brief explanation is given in Appendix~\ref{sec:evalScalarFunctions}. The transverse scalar \(y-\phi\) correlation is again no longer constant, and evolves as \(\e\) changes value. Similarly, the angular power spectra for \(\e<0\) can be computed only numerically using the same method as earlier, and are plotted in Fig.~\ref{fig:superluminalScalarCoeffs}. 

If \textsc{gw}s were scalar transverse instead of tensorial, the astrometric deflection correlation curve would change from that plotted in Fig.~\ref{fig:subluminalTensorialCorr} to that plotted in Fig.~\ref{fig:subluminalScalarCorr} (blue curves in both figures). This change could equivalently be explained through the corresponding angular power spectra. The correlation spectrum would change from being predominantly quadrupolar for a tensorial \textsc{gw} background to a dipolar (\(\ell=1\)) for a scalar background; see the blue curves in Figs.~\ref{fig:subluminalTensorialCoeffs} and \ref{fig:subluminalScalarCoeffs}). Note that the angular power spectrum for a scalar background is strictly unimodal (i.e., only has a single non-zero angular mode) when \(\e=0\).

Additionally, if \textsc{gw}s were to propagate at a speed lower or larger than the speed of light then the correlation pattern would generally fail to match the previously established predictions. The effects of \(\e \neq 0\) are best explained through Figs.~\ref{fig:subluminalScalarCoeffs} and \ref{fig:superluminalScalarCoeffs}. The luminal case is always unimodal, however non-luminal \textsc{gw}s would induce angular power at higher multipole orders. As the magnitude of \(\e\) is increased, this new power distribution concentrates power at lower-order modes for sub-luminal \textsc{gw}s, and increases power at higher-order modes for super-luminal \textsc{gw}s.

\subsection{Vectorial Polarizations}
\label{sec:vectorial}
\noindent
Some modified theories of gravity may also contain \textsc{gw} polarizations which have a mixture of transverse and longitudinal components. The two polarization states of such nature are described as ``vectorial'' modes (because they transform with spin weight 1 under rotations around the \textsc{gw} wavevector), and their effect has to be considered together, as their signatures are rotational variants of each other. In this subsection are considered  backgrounds that consist solely of vectorially polarized \textsc{gw}s, i.e. the parameter \(P\) runs over the set \(\{X, Y\}\).

The sub-luminal correlation functions can be found analytically in an analogous way to before. Details of the derivation are given in Appendix~\ref{sec:evalVectorialFunctions} and the results are given by eqs.~(\ref{eq:SublimunalVectorialxthCorr}) and (\ref{eq:SublimunalVectorialyphCorr}). These are plotted for four representative values of \(\e\) in Fig.~\ref{fig:subluminalVectorialCorr}. The angular power spectra for these values of \(\e\) are shown in Fig.~\ref{fig:subluminalVectorialCoeffs}, with the luminal case considered in further detail in Appendix~\ref{app:vectorialLuminalSpectra}.

Similar to the previous cases of interest, here the super-luminal correlation functions cannot be computed analytically, but only numerically; they are presented in Fig.~\ref{fig:superluminalVectorialCorr} and some brief notes on calculating them are provided in Appendix~\ref{sec:evalVectorialFunctions}. The angular power spectra for \(\e<0\) can likewise be computed using the same numerical method as before, and are presented in Fig.~\ref{fig:superluminalVectorialCoeffs}.

For a background of vectorially-polarized \textsc{gw}s, the astrometric deflection correlation curve would no longer look like the one plotted in Fig.~\ref{fig:subluminalTensorialCorr}, but would change to that presented in Fig.~\ref{fig:subluminalVectorialCorr} (blue curves in both figures). From a spectral perspective, the angular power spectrum of the correlation would shift from being predominantly quadrupolar for tensorial modes to sharing most of the spectral power between the dipole and the quadrupole (\(\ell=1, 2\)) in the vectorial case; see blue curves in Figs.~\ref{fig:subluminalTensorialCoeffs} and \ref{fig:subluminalVectorialCoeffs}).

Moreover, if the \textsc{gw}s were to propagate at a speed lower (larger) than the speed of light then the correlation pattern would again no longer follow the prediction for \(\e=0\) in Fig.~\ref{fig:subluminalVectorialCorr}. In this case the change would mean a reduction (enhancement) of the spectral power at higher multipole orders (see Figs.~\ref{fig:subluminalVectorialCoeffs} and \ref{fig:superluminalVectorialCoeffs}), compared to the \(\e = 0\) case.

\subsection{Scalar Longitudinal Polarization}
\label{sec:longitudinal}
\noindent
The \(L\) polarization state is a purely longitudinal, scalar mode. For a massive theory of gravity, its effect is indistinguishable from the effect of the transverse scalar mode. In this subsection is considered a stochastic background of purely \(L\)-polarized \textsc{gw}s. In the limit \(d \rightarrow \infty\) the astrometric response \(\delta n_{i}^{\mathsc{L}} (\vect{n}, \vect{q}, d)\) is divergent for \(\vect{n} \parallel \vect{q}\), however this is alleviated by including a positive \(\e\) in the equation.

The sub-luminal correlation functions can be obtained analytically in an analogous way to before. Details of their derivation are presented in Appendix~\ref{sec:evalLongitudinalFunctions} and the final results are given by eqs.~(\ref{eq:SublimunalLongitudinalxthCorr}) and (\ref{eq:SublimunalLongitudinalyphCorr}). The correlation curves are plotted for four representative values of \(\e\) in Fig.~\ref{fig:subluminalLongitudinalCorr}. The sub-luminal angular power spectra for the same values of \(\e\) are shown in Fig.~\ref{fig:subluminalLongitudinalCoeffs}, with the luminal case considered in additional detail in Appendix~\ref{app:longitudinalLuminalSpectra}.

The super-luminal correlation functions cannot be derived analytically; they can only be computed numerically; these are plotted in Fig.~\ref{fig:superluminalLongitudinalCorr} and brief notes on calculating them are supplied in Appendix~\ref{sec:evalLongitudinalFunctions}. The angular power spectra for \(\e<0\) can be computed by numerically integrating eqs.~(\ref{eq:spectraCoeffsFormulae}), and are plotted in Fig.~\ref{fig:superluminalLongitudinalCoeffs}.

If the stochastic background was comprised of scalar longitudinal \textsc{gw}s, the astrometric deflection correlation curve would have the shape of the one plotted in Fig.~\ref{fig:subluminalLongitudinalCorr} (blue curve). Additionally, the angular spectrum of the correlation would appear to be predominantly quadrupole in this case, with smaller dipolar and octopolar contributions (\(\ell=1, 2, 3\); see blue curve in Fig.~\ref{fig:subluminalLongitudinalCoeffs}).

Finally, if these \textsc{gw}s were to propagate with a non-luminal group velocity, the astrometric correlation pattern would again differ from the \(\e = 0\) prediction in Fig.~\ref{fig:subluminalLongitudinalCorr}. For sub-(super-)luminal velocities, this change would induce a reduction (enhancement) of the spectral power at higher multipole orders (see Figs.~\ref{fig:subluminalLongitudinalCoeffs} and \ref{fig:superluminalLongitudinalCoeffs}), compared to the \(\e = 0\) case.

\section{Connection to pulsar timing methods}
\label{sec:additionalResults}
\noindent
In the previous sections of the article, analytical results for the sub-luminal correlation functions of astrometric deflections due to a stochastic \textsc{gw} background were presented. However, these were not the only correlations which are of interest in the context of \textsc{gw} backgrounds. The correlations of redshift of millisecond pulsars are in the core of the \textsc{pta}-based method to probe for such backgrounds, and in \citep{2018PhRvD..97l4058M} the authors presented the case for cross-correlating the redshift of a pulsar with the astrometric deflection of a star, opening up the new possibility of combined pulsar timing and astrometric \textsc{gw} searches. In all previous literature, the \textsc{pta} correlation functions for a background of sub-luminal \textsc{gw}s have been explored only numerically. In this section, analytic results for both types of correlations are presented. The derivations are performed in a similar way to the results presented in Sections~\ref{sec:tensorial} -- \ref{sec:longitudinal}.

\subsection{Redshift correlations}
\label{sec:redshiftResults}
\noindent
The redshift correlations arising from a background of sub-luminal \textsc{gw}s has been studied previously in the context of pulsar timing \citep{Baskaran:2008za, PhysRevD.78.089901, Lee:2010cg, Lee:2014awa, Yunes:2013dva, Gumrukcuoglu:2012wt}. The dependence of the correlation curves on \(\e\) has been investigated and discussed before, notably in \citep{Lee:2010cg}. In their calculation the correlation integral was evaluated numerically -- that was necessary because \citep{Lee:2010cg} considered a stochastic background of \textsc{gw}s with a power-law spectrum in frequency. Each frequency component of the background is a stochastic process and propagates with its own speed parameter \(\e\) given by eq.~(\ref{eq:edef}). Integrating over the entire frequency spectrum to calculate the resultant correlation curve may only be performed numerically. In this paper the background is assumed to be characterised by a single speed parameter; this is equivalent to assuming that the background only has power in a narrow frequency band \(\Delta f\) which is small compared to the reciprocal of the observation duration \(T\). The results of this calculation are shown in Fig.~\ref{fig:redshiftCorrelation} and Appendix~\ref{sec:AppRedshift}; in particular, it is interesting to compare the top left panel of this figure with Fig.~2 of \citep{Lee:2010cg} which shows that the two approaches give results which are in excellent agreement.

The \textsc{pta}-\textsc{pta} correlation is defined through the following integral, in analogy to the definition in eqs.~(\ref{eq:GammaDef})
\begin{align}\label{eq:GammazzDef}
\Gamma^{\mathsc{P}}_{zz}(\Theta) = \int_{S^{2}} \!\!\!\dd\Omega_{\vect{q}}\; z^{\mathsc{P}}(\vect{n}, \vect{q})\,z^{\mathsc{P}}(\vect{m}, \vect{q})\,,
\end{align}
where \(z^{\mathsc{P}}(\vect{n}, \vect{q})\) is the redshift of a pulsar at \(\vect{n}\) due to a \textsc{gw} coming from \(\vect{q}\) in a background of \(P\)-polarized \textsc{gw}s,
\begin{align}\label{eq:redshiftDef}
z^{\mathsc{P}}\!(\vect{n}, \vect{q}, d, t) = \frac{n^{i}n^{j}}{2(1 - q^{\ell} n_{\ell})}\Big(\!1 - \exp\!\big(\!-\!\rmi d (1 - q^{s} n_{s})\big)\!\Big)h_{ij} (t).
\end{align}
The temporal dependence in eq.~(\ref{eq:GammazzDef}) has been factored out analogously to eq.~(\ref{eq:new1}). While in the astrometric case the star term was commonly discarded by assuming the distant-source limit, here this is not possible as the Earth and pulsar (the equivalent of the star term in eq.~(\ref{eq:astrometricShiftFull}), i.e., the value of the \textsc{gw} perturbation at the pulsar) terms have the same relative weight (see the brackets in eq.~(\ref{eq:redshiftDef})). However, since the distances from Earth to the pulsars on the sky are essentially uncorrelated, the pulsar term in eq.~(\ref{eq:redshiftDef}) can be left out when doing the correlation integral. The well-known Hellings-Downs curve is derived from eq.~(\ref{eq:GammazzDef}) by restricting \(P \in \{+, \times\}\) and \(\e = 0\) \citep{1983ApJ...265L..39H}. Thus, the integral in eq.~(\ref{eq:GammazzDef}) can be solved analytically for each separate polarization content. The results are given in Appendix~\ref{sec:AppRedshift}.

\subsection{Redshift-astrometry correlations}
\label{sec:redshiftAstrometryResults}
\noindent
In \citep{2018PhRvD..97l4058M} the authors presented correlation functions between redshift and astrometry measurements, for a stochastic background of luminal \textsc{gw}s. Similar to the other two cases already presented in the current article, that of astrometry- astrometry and redshift-redshift correlation, it is possible to expand on these results by deriving the corresponding functions for \(\e > 0\), defined by (cf. eqs.~(\ref{eq:GammaDef}) and (\ref{eq:GammazzDef}))
\begin{align}\label{eq:GammazbDef}
\Gamma^{\mathsc{P}}_{zb}(\Theta) = \int_{S^{2}} \!\!\!\dd\Omega_{\vect{q}}\; z^{\mathsc{P}}(\vect{n}, \vect{q})\, \delta m_{b}^{\mathsc{P}}(\vect{q})\,,
\end{align}
where the pulsar term in the redshift and the star term in the astrometric deflection are ignored. This is the cross correlation between the redshift in direction \(\vect{n}\) and the two components (\(b\in(\theta, \phi)\)) of the astrometric deflection in direction \(\vect{m}\). The results are shown in Fig.~\ref{fig:redshiftAstrometryCorrelation} in Appendix~\ref{sec:AppRedshiftAstrometry}.

\section{Discussion and conclusions}
\label{sec:discussion}
\noindent
A \textsc{gw} produces small perturbations in the apparent position (observable as an astrometric deflection) and redshift (observable as an integrated timing residual) of a distant object on the sky. Furthermore, an isotropic, stochastic background of \textsc{gw}s produces a correlated, stochastic pattern of redshifts and of astrometric deflections over the sphere of the sky. Within the theory of general relativity \textsc{gw}s exist in only two polarization states and propagate in vacuum at the speed of light. If additionally it is assumed that the \textsc{gw} background is statistically stationary, isotropic, unpolarised, etc., then \textsc{gr} makes a unique prediction for these correlation patterns -- the prediction contains no free parameters. The \textsc{gr} prediction for the redshift correlation between two points on the sky separated by an angle \(\Theta\) is known as the Hellings-Downs function \cite{1983ApJ...265L..39H}. Similarly, the correlation between the components of the astrometric deflections at two points is also completely described by a single function of \(\Theta\), the astrometric analog of the \textsc{hd} curve (see the blue curve in Fig.~\ref{fig:subluminalTensorialCorr} and \citep{PhysRevD.83.024024,2018PhRvD..97l4058M, 2018PhRvD..98b4020O}).

Instead of describing the correlations as functions of the angular separation, it is convenient to decompose the redshift and astrometric deflection (scalar and vector fields over the sphere) in terms of (vector) spherical harmonics and to study the correlations between the coefficients of this expansion. Translated into this description, the unique \textsc{gr} prediction for these correlations is that they are predominantly quadrupolar, meaning the dominant correlation is between the \(\ell = 2\) modes (see, for example, Fig.~\ref{fig:subluminalTensorialCoeffs}). These two methods of description, correlation curve or spectral structure, are completely equivalent; Sec.~\ref{sec:theory} of this paper describes how to relate the two descriptions.

While the theory of general relativity predicts only 2 polarization modes, propagating at the speed of light, modified metric theories of gravity can predict up to 4 additional \textsc{gw} polarizations propagating at either sub- or super-luminal speeds. Recent observations of \textsc{gw}s with frequency of \(\sim \SI{E2}{\hertz}\) with \textsc{ligo} and Virgo have placed constraints on the speed \cite{2017PhRvL.118v1101A} and polarization content \cite{Abbott:2018lct, Abbott:2017oio} of \textsc{gw}s; future results from Earth-based detectors will further improve these constraints. Pulsar timing and astrometric measurements offer the possibility to test the propagation of \textsc{gw}s with considerably lower frequency. As discussed in the introduction of this article, if gravitons are dispersed in vacuum like massive particles then the \textsc{gw} group velocity is controlled by the ratio \((m_{\mathrm{g}} / \omega)^{2}\), so the effect of the graviton mass is exaggerated at lower frequencies. The advent of low-frequency \textsc{gw} astronomy would make possible for this effect to be examined from a fresh perspective, and place a new model-independent constraint on the value of the speed of \textsc{gw}s. This work extends the earlier calculations in \citep{PhysRevD.83.024024}, \citep{2018PhRvD..97l4058M} and \cite{2018PhRvD..98b4020O} by calculating the expected redshift and astrometric deflection correlation patterns generated by stochastic \textsc{gw} backgrounds with arbitrary polarizations and propagating with arbitrary speeds.

Section~\ref{sec:non-luminalCorrelations} presents the theoretical predictions for the correlation patterns for each polarization content. The standard \textsc{gr} case is covered in Subsection~\ref{sec:tensorial}: the astrometric correlation function is given by the blue curve in Fig.~\ref{fig:subluminalTensorialCorr}. In addition, the same subsection covers the case of the tensorial modes propagating at a non-luminal group velocity, see Figs.~\ref{fig:subluminalTensorialCorr} and \ref{fig:superluminalTensorialCorr} for the sub- and super-luminal correlation curves, respectively. Furthermore, the same results are presented from the standpoint of angular power spectra: Figs.~\ref{fig:subluminalTensorialCoeffs} and \ref{fig:superluminalTensorialCoeffs} show the spectral distributions in the sub- and super-luminal regimes, and the luminal case for comparison. The same analysis is then repeated for the additional polarization contents: Subsections~\ref{sec:scalar}, \ref{sec:vectorial}, and \ref{sec:longitudinal} explore the cases of scalar transverse, vectorial, and scalar longitudinal stochastic \textsc{gw} backgrounds. In each subsection are presented the findings for the shape of the correlation curves, in the luminal, sub-luminal, and super-luminal cases, plus the angular power spectra distributions which correspond to each correlation function.

The results in Section~\ref{sec:non-luminalCorrelations} reveal the effect of a non-luminal group velocity of the \textsc{gw}s in a stochastic background: in general, sub-luminal speeds enhance the power in low-order modes, while super-luminal speeds increase the power in high-order modes. Some cases (e.g. the scalar transverse polarization, see Subsection~\ref{sec:scalar}) are more involved, but the general trend holds everywhere.

The results presented in Section~\ref{sec:non-luminalCorrelations} can be summarised in the following manner: the astrometric two-point correlation function on the sky depends on the polarization content and the propagation speed of the stochastic gravitational wave background. The formalism of angular power distribution makes it more straightforward to compare these results with future observations, from \emph{Gaia} or other astrometric catalogues. 
Assuming such a measurement can be made, then the observed correlations should naturally be compared against the \textsc{gr} predictions. Should a deviation be found then the results of this paper would help to relate the observed deviation to a theoretical description of the type of departure from \textsc{gr}. Alternatively, if no deviation is found within the experimental uncertainties, then the results in this paper will be necessary when attempting to convert the observed correlations into constraints on the possible deviations from \textsc{gr}.

Section~\ref{sec:additionalResults} presents some additional results which would be of interest to the reader: Subsection~\ref{sec:redshiftResults} revisits the case of pulsar timing correlations. Subsection~\ref{sec:redshiftAstrometryResults}, on the other hand, investigates the cross-correlation between the timing residual of a pulsar, and the astrometric deflection of a star, first introduced in \citep{2018PhRvD..97l4058M}. In each case, the correlation formalism is introduced and explicit expressions are calculated for the sub-luminal regime. These results would provide methods for testing the predictions of the astrometric correlations from Section~\ref{sec:non-luminalCorrelations}.

Future developments on this topic would include incorporating these theoretical predictions into a numerical analysis tool to test the upcoming \emph{Gaia} Data Releases. Further work could include testing the anisotropy of the background, or placing constraints on the stochastic background using the proper motions of quasi-stellar objects.

\vspace*{-10pt}

\section*{Acknowledgements}
\noindent
DM acknowledges support from the EU FP7 programme through ERC Grant no. 320360 (University of Cambridge) as well as support from the UK Space Agency, grant reference ST/R001901/1 (University of Edinburgh).

\bibliographystyle{apsrev4-1.bst}
\bibliography{main}

\onecolumngrid
\clearpage

\begin{appendices}
\numberwithin{equation}{section}
\renewcommand{\theequation}{\Alph{section}\arabic{equation}}
\renewcommand{\thesubsection}{\Roman{subsection}}

\section{The Astrometric Shift}
\label{sec:AstrometricShift}
\begin{figure}[b]
\includegraphics[scale=1]{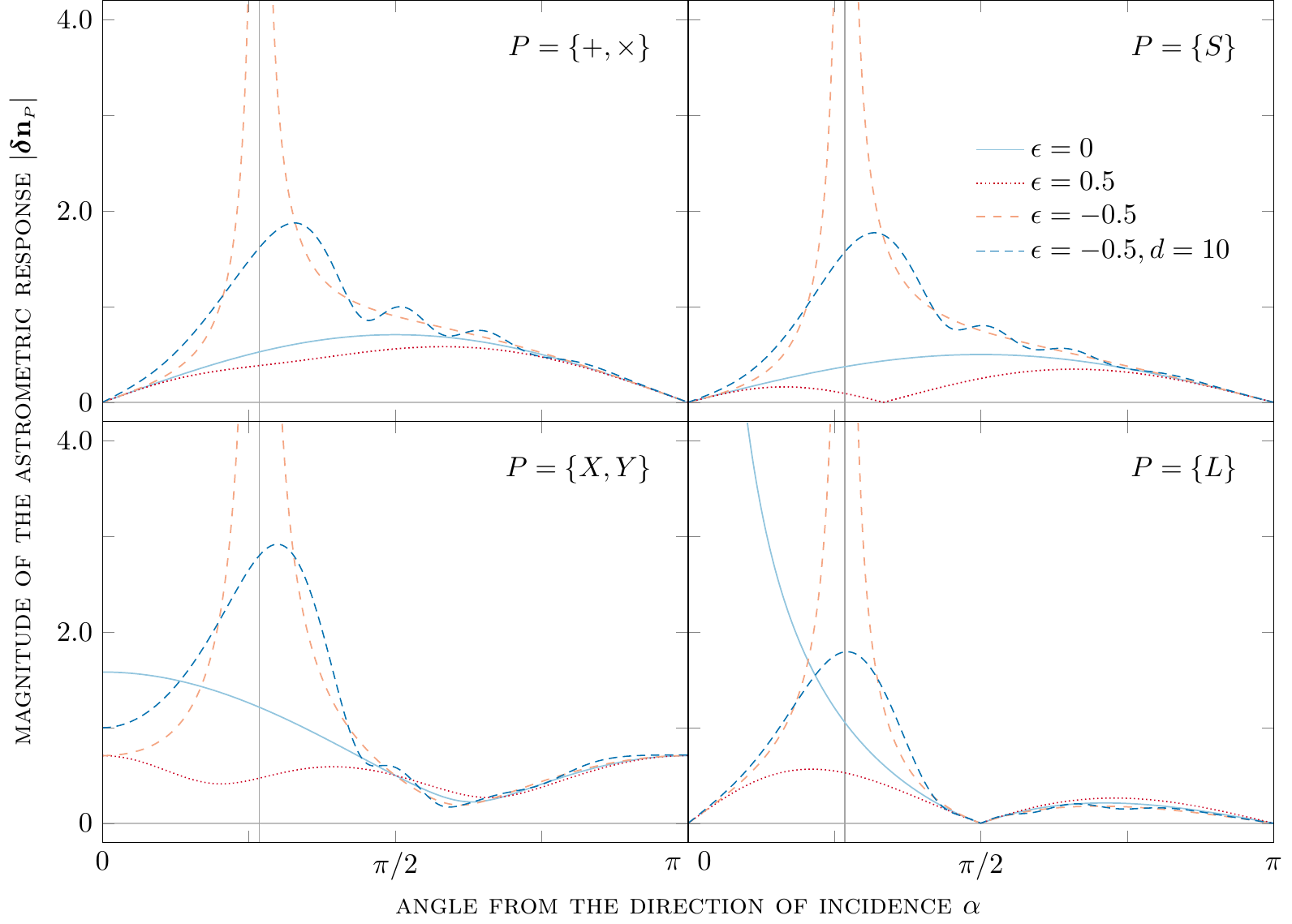}
\caption{Relative amplitude of the astrometric shift as a function of the angle from the direction \(\vect{q}\) to the \textsc{gw} source, for all possible polarization contents and for different regimes of the parameter \(\e\). In the distant-source limit, for \(\e = 0\) the astrometric shift is non-divergent in all cases expect for \(P = \{L\}\). A positive value of \(\e = 0.5\) removes the divergence in this case. For negative \(\e = -0.5\) the astrometric shift is always divergent in the distant-source limit, at an angle of \(\alpha = \arcsec (1 - \e)\) (marked with a vertical line on the plots). This divergence is unphysical and can be removed by including the star term in the astrometric shift formula, here demonstrated with \(d = 10\) gravitational wavelengths.}
\label{fig:astrometricShift}
\end{figure}

\noindent
The astrometric shift, specified in Section~\ref{sec:theory} has been used throughout this article. To better illustrate its properties, it is useful to show how it depends on its parameters. In Fig.~\ref{fig:astrometricShift} are shown plots of the magnitude of the astrometric shift \(|\vect{\delta n}_{\mathsc{P}}|\) for all 4 possible polarization contents. The plot serves to demonstrate what has already been explained through equations in the man text -- the long-distance limit of the astrometric shift, defined through eq.~(\ref{eq:astrometricShiftLimit}), is unphysically divergent for all polarizations at \(\alpha = \arcsec (1 - \e)\) when \(\e < 0\), and in the scalar longitudinal case at \(\alpha = 0\) when \(\e = 0\). The full formula for the astrometric shift, specified by eq.~(\ref{eq:astrometricShiftFull}) is regular for all polarization contents, and for all values of \(\e\) and \(\alpha\), although it peaks strongly around the angle of the cone on which the distant source expression diverges.

\section{The astrometric correlation integrals and results}
\label{app:evalFunctions}
\noindent
In Sections~\ref{sec:tensorial} to \ref{sec:longitudinal}, the details of the evaluations of the spatial correlation integrals for the different polarization modes were omitted for brevity; these details and corresponding results are presented in this Appendix.

\subsection{Tensorial transverse-traceless correlations}
\label{sec:evalTensorialFunctions}
\noindent
In Section~\ref{sec:theory} of the main body of the text, it was explained why the spatial correlation matrix in eq.~(\ref{eq:corrMatrix}) is strictly 2-dimensional. The non-zero components of this correlation matrix are specified by two scalar integrals over the sky; one involving \(x\) and \(\theta\) components, and one involving \(y\) and \(\phi\) components. In this Appendix the method for evaluating these two integrals will be shown, first for the \(+\) mode and then for the \(\times\) mode.

Firstly, the \(x-\theta\) and \(y-\phi\) correlation integrals for the \(+\) polarized \textsc{gw} state are considered; the correlation integrals are defined in eqs.~(\ref{eq:new2}) and (\ref{eq:new3}) as
\begin{subequations}\label{eq:plusIntegralEval}
\begin{align}
\Gamma^{+}_{x\theta}(\Theta, \e) = \int_{S^{2}} \!\!\!\dd\Omega_{\vect{q}}\;\delta n_{x}^{+}(\vect{q})\,\delta m_{\theta}^{+}(\vect{q}) = \int_{0}^{2\pi} \!\!\!\!\dd\phi \int_{0}^{\pi} \!\!\!\dd\theta\,\sin\theta\,\delta n_{x}^{+}(\e, \theta, \phi)\,\delta m_{\theta}^{+}(\e, \theta, \phi) \,, \\
\Gamma^{+}_{y\phi}(\Theta, \e) = \int_{S^{2}} \!\!\!\dd\Omega_{\vect{q}}\;\delta n_{y}^{+}(\vect{q})\,\delta m_{\phi}^{+}(\vect{q}) = \int_{0}^{2\pi} \!\!\!\!\dd\phi \int_{0}^{\pi} \!\!\!\dd\theta\,\sin\theta\,\delta n_{y}^{+}(\e, \theta, \phi)\,\delta m_{\phi}^{+}(\e, \theta, \phi) \,,
\end{align}
\end{subequations}
where the vector \(\vect{q} = (1 - \e) (\sin\theta \cos\phi, \sin\theta \sin\phi, \cos\theta)\) is the direction on the sky from which the \textsc{gw} originates. The components of the astrometric deflection, \(\delta n_{x}^{+}(\e, \theta, \phi)\), \(\delta m_{y}^{+}(\e, \theta, \phi)\), \(\delta n_{\theta}^{+}(\e, \theta, \phi)\), and \(\delta m_{\phi}^{+}(\e, \theta, \phi)\), may be evaluated using the formula in eq.~(\ref{eq:astrometricShiftDef}), the distant-source limit in eq.~(\ref{eq:astrometricShiftLimit}), and the prescription described in Appendix B of \citep{2018PhRvD..97l4058M}. Using \(F_{ab}^{+}\) to denote the aggregated integrand, eqs.~(\ref{eq:plusIntegralEval}) take the form

\begin{subequations}\label{eq:plusIntegralEvalSuper}
\noindent\begin{minipage}[l]{0.555\linewidth}
\begin{align}
\Gamma^{+}_{x\theta}(\Theta, \e) = \int_{0}^{2\pi} \!\!\!\!\dd\phi \int_{0}^{\pi} \!\!\!\dd\theta\,F_{x\theta}^{+}(\Theta, \e, \theta, \phi)\,,
\end{align}
\end{minipage}\hspace*{0.03\linewidth}
\begin{minipage}[l]{0.41\linewidth}
\begin{align}
\Gamma^{+}_{y\phi}(\Theta, \e) = \int_{0}^{2\pi} \!\!\!\!\dd\phi \int_{0}^{\pi} \!\!\!\dd\theta\,F_{y\phi}^{+}(\Theta, \e, \theta, \phi)\,,
\end{align}
\end{minipage}
\end{subequations}\\[7pt]
where the \(\Theta\) dependence originates from the orientation of the vector \(\vect{\hat{m}}\) relative to the vector \(\vect{\hat{n}}\). For \(\e > 0\), these two double integrals may now be evaluated by applying consecutively the techniques described in Appendices~\ref{app:azimuthalIntegral} and \ref{app:polarIntegral} to obtain functional forms for \(\Gamma^{+}_{x\theta}(\Theta, \e)\) and \(\Gamma^{+}_{y\phi}(\Theta, \e)\).

Secondly, the \(x-\theta\) and \(y-\phi\) correlation integrals for the \(\times\) polarized \textsc{gw} state are considered; the correlation integrals are defined in eqs.~(\ref{eq:GammaDef}), the vector \(\vect{q}\) is the direction on the sky from which the \textsc{gw} originates, and the components of the astrometric deflection may be evaluated in analogous way to before. Therefore, these take the form
\begin{subequations}\label{eq:crossIntegralEval}
\begin{align}
\Gamma_{x\theta}^{\times} (\Theta, \e) = \int_{S^{2}} \!\!\!\dd\Omega_{\vect{q}}\;\delta n_{x}^{\times}(\vect{q})\,\delta m_{\theta}^{\times}(\vect{q}) = \int_{0}^{2\pi} \!\!\!\!\dd\phi \int_{0}^{\pi} \!\!\!\dd\theta\,\sin\theta\,\delta n_{x}^{\times}(\e, \theta, \phi)\,\delta m_{\theta}^{\times}(\e, \theta, \phi) = \int_{0}^{2\pi} \!\!\!\!\dd\phi \int_{0}^{\pi} \!\!\!\dd\theta\,F_{x\theta}^{\times}(\Theta, \e, \theta, \phi) \,, \\
\Gamma_{y\phi}^{\times} (\Theta, \e) = \int_{S^{2}} \!\!\!\dd\Omega_{\vect{q}}\;\delta n_{y}^{\times}(\vect{q})\,\delta m_{\phi}^{\times}(\vect{q}) = \int_{0}^{2\pi} \!\!\!\!\dd\phi \int_{0}^{\pi} \!\!\!\dd\theta\,\sin\theta\,\delta n_{y}^{\times}(\e, \theta, \phi)\,\delta m_{\phi}^{\times}(\e, \theta, \phi) = \int_{0}^{2\pi} \!\!\!\!\dd\phi \int_{0}^{\pi} \!\!\!\dd\theta\,F_{y\phi}^{\times}(\Theta, \e, \theta, \phi) \,.
\end{align}
\end{subequations}
For \(\e > 0\), these two double integrals may now be evaluated by applying consecutively the techniques described in Appendices~\ref{app:azimuthalIntegral} and \ref{app:polarIntegral} to obtain functional forms for \(\Gamma^{\times}_{x\theta}(\Theta, \e)\) and \(\Gamma^{\times}_{y\phi}(\Theta, \e)\). For an unpolarised background containing equal power of both \(+\) and \(\times\) polarization states the combined spatial correlation functions may be defined as \(\Gamma^{+, \times}_{x\theta}(\Theta, \e) = \Gamma^{+}_{x\theta}(\Theta, \e) + \Gamma^{\times}_{x\theta}(\Theta, \e)\) and \(\Gamma^{+, \times}_{y\phi} (\Theta, \e) = \Gamma^{+}_{y\phi} (\Theta, \e) + \Gamma^{\times}_{y\phi} (\Theta, \e)\). These new functions may be evaluated by taking the sums of the expression derived above to give the result which appeared in the main text, eqs.~(\ref{eq:subluminalTensorialxthCorr}) and (\ref{eq:subluminalTensorialyphCorr}).

For \(\e < 0\), the double integrals in eqs.~(\ref{eq:plusIntegralEvalSuper}) and (\ref{eq:crossIntegralEval}) could not be evaluated as the astrometric response is divergent at \(\Theta = \arcsec (1 - \e)\) (see Appendix~\ref{sec:AstrometricShift}). In order to find a meaningful interpretation of the correlation function \(\G_{ab}^{+, \times}\), one has to include the star terms in the astrometric response at each star, i.e. use eq.~(\ref{eq:astrometricShiftDef}) with eq.~(\ref{eq:astrometricShiftFull}). The resulting integrals would also depend on the distance to the star in units of \textsc{gw} wavelengths, \(d\):
\begin{subequations}\label{eq:plusCrossIntegralEval}
\begin{align}
\Gamma_{x\theta}^{+, \times} (\Theta, \e, d) = \int_{S^{2}} \!\!\!\dd\Omega_{\vect{q}}\;\delta n_{x}^{+, \times}(\vect{q}, d)\,\delta m_{\theta}^{+, \times}(\vect{q}, d) = \int_{0}^{2\pi} \!\!\!\!\dd\phi \int_{0}^{\pi} \!\!\!\dd\theta\,F_{x\theta}^{+, \times}(\Theta, \e, d, \theta, \phi) \,, \\
\Gamma_{y\phi}^{+, \times} (\Theta, \e, d) = \int_{S^{2}} \!\!\!\dd\Omega_{\vect{q}}\;\delta n_{y}^{+, \times}(\vect{q}, d)\,\delta m_{\phi}^{+, \times}(\vect{q}, d) = \int_{0}^{2\pi} \!\!\!\!\dd\phi \int_{0}^{\pi} \!\!\!\dd\theta\,F_{y\phi}^{+, \times}(\Theta, \e, d, \theta, \phi) \,.
\end{align}
\end{subequations}
These integrals, however, were not found to have analytical solutions and could only be evaluated numerically. Numerical integration (for a specific \(\e\) and \(d\)) yields the curves shown in Fig.~\ref{fig:superluminalTensorialCorr}.

\subsubsection{Azimuthal integral}
\label{app:azimuthalIntegral}
\noindent
The following integral appears in all spatial correlation integrals in this article:
\begin{align}
J_{n} (\theta, \Theta, \e) = \!\int_{0}^{2\pi} \!\!\!\!\! \dd\phi\,\frac{\cos(n\phi)}{1 - (1-\e)\cos\Theta \cos\theta - (1-\e)\sin\Theta \sin\theta \cos\phi}.
\end{align}
It can be solved by referring to result (B17a) in Appendix~A.I of \citep{2018PhRvD..97l4058M}:
\begin{align}\label{eq:azimuthalIntSol}
\begin{split}
&J_{n} (\theta, \Theta, \e) = \frac{2\pi}{\sqrt{(1 - (1-\e)\cos\Theta \sin\theta)^{2} - \sin^{2}\!\Theta \sin^{2}\!\theta}}\,\times \\
&\RepQuad{12} \times \left(\frac{1 - (1-\e)\cos\Theta \sin\theta - \sqrt{(1 - (1-\e)\cos\Theta \sin\theta)^{2} - \sin^{2}\!\Theta \sin^{2}\!\theta}}{(1-\e)\sin\Theta \sin\theta}\vphantom{\left(\Bigg[\Bigg]^{2}\right)^{2}}\right)\np{n}.
\end{split}
\end{align}

\subsubsection{Polar integral}
\label{app:polarIntegral}
\noindent
The polar integrals which come up in the derivation of the sub-luminal correlation functions (see Section~\ref{sec:theory}) all have the following form
\begin{align}\label{eq:polarInt}
K_{n} (\Theta, \e) = \!\int_{0}^{\pi} \!\!\! \dd\theta\,\sin^{n}(\theta)\,J_{m} (\theta, \Theta, \e),
\end{align}
where \(J_{m} (\theta, \Theta, \e)\) is defined in Appendix~\ref{app:azimuthalIntegral}, \((l + m)\) is an even number, which guarantees that the overall \(\sin(\theta)\) factor has even power. This integral can be most readily addressed using the substitution \(\sqrt{2\e - \e^{2}}\,\sin(\Theta) \sinh(u) = (1-\e) \sin(\theta) - \cos(\Theta)\), using which the square root in eq.~(\ref{eq:azimuthalIntSol}) becomes
\begin{align}
\sqrt{(1 - (1-\e)\cos\Theta \sin\theta)^{2} - \sin^{2}\!\Theta \sin^{2}\!\theta}\,= \sqrt{2\e - \e^{2}}\,\sin(\Theta) \cosh(u),
\end{align}
and the integrand of (\ref{eq:polarInt}) is now a rational function of hyperbolic functions of \(u\). Finally, using the substitution \(v = e^{u}\), the integrand becomes a rational function of \(v\), which can then be analytically integrated.

\subsection{Scalar transverse correlations}
\label{sec:evalScalarFunctions}
\noindent
In this Appendix the evaluation of spatial correlation integrals for the transverse scalar \textsc{gw} polarization state, \(S\), is briefly described. The \(x-\theta\) and \(y-\phi\) correlation integrals are defined in eqs.~(\ref{eq:GammaDef}), the vector \(\vect{q}\) is the direction on the sky from which the \textsc{gw} originates, and the components of the astrometric deflection may be evaluated in an analogous way to the one described in Appendix~\ref{sec:evalTensorialFunctions}. Therefore, these take the form
\begin{subequations}\label{eq:scalarIntegralEval}
\begin{align}
\Gamma_{x\theta}^{\mathsc{S}} (\Theta, \e) = \int_{S^{2}} \!\!\!\dd\Omega_{\vect{q}}\;\delta n_{x}^{\mathsc{S}}(\vect{q})\,\delta m_{\theta}^{\mathsc{S}}(\vect{q}) = \int_{0}^{2\pi} \!\!\!\!\dd\phi \int_{0}^{\pi} \!\!\!\dd\theta\,\sin\theta\,\delta n_{x}^{\mathsc{S}}(\e, \theta, \phi)\,\delta m_{\theta}^{\mathsc{S}}(\e, \theta, \phi) = \int_{0}^{2\pi} \!\!\!\!\dd\phi \int_{0}^{\pi} \!\!\!\dd\theta\,F_{x\theta}^{\mathsc{S}}(\Theta, \e, \theta, \phi) \,, \\
\Gamma_{y\phi}^{\mathsc{S}} (\Theta, \e) = \int_{S^{2}} \!\!\!\dd\Omega_{\vect{q}}\;\delta n_{y}^{\mathsc{S}}(\vect{q})\,\delta m_{\phi}^{\mathsc{S}}(\vect{q}) = \int_{0}^{2\pi} \!\!\!\!\dd\phi \int_{0}^{\pi} \!\!\!\dd\theta\,\sin\theta\,\delta n_{y}^{\mathsc{S}}(\e, \theta, \phi)\,\delta m_{\phi}^{\mathsc{S}}(\e, \theta, \phi) = \int_{0}^{2\pi} \!\!\!\!\dd\phi \int_{0}^{\pi} \!\!\!\dd\theta\,F_{y\phi}^{\mathsc{S}}(\Theta, \e, \theta, \phi) \,.
\end{align}
\end{subequations}
For \(\e > 0\), these two double integrals may now be evaluated by applying consecutively the techniques described in Appendices~\ref{app:azimuthalIntegral} and \ref{app:polarIntegral} to obtain functional forms for \(\Gamma^{\mathsc{S}}_{x\theta} (\Theta, \e)\) and \(\Gamma^{\mathsc{S}}_{y\phi} (\Theta, \e)\):
\begin{subequations}
\begin{align}
&\G_{x\theta}^{\mathsc{S}} (\Theta, \e) = \frac{\pi}{3}\,\frac{\left(1-8\e-8\e^{2}+12\e^{3}-3\e^{4}\right) - 2\!\left(1-8\e+4\e^{2}\right)\!\s[2]}{(1-\e)^{4}} + \frac{\pi}{2}\,\frac{\e^{2}(2-\e)^{2}\!\left(2-3\s[2]+2\s[4]\right)}{(1-\e)^{5}\!\left(1-\s[2]\right)}\,\eln \nonumber\\
&\RepQuad{5}- \pi\,\frac{\e^{2}(2-\e)^{2}\s}{(1-\e)^{5}\!\left(1-\s[2]\right)\!\ssqrt}\,\sln , \label{eq:SublimunalScalarxthCorr}\\[10pt]
\begin{split}\label{eq:SublimunalScalaryphCorr}
&\G_{y\phi}^{\mathsc{S}} (\Theta, \e) = \frac{\pi}{3}\,\frac{1-8\e+4\e^{2}}{(1-\e)^{4}} + \frac{\pi}{2}\,\frac{\e^{2}(2-\e)^{2}\!\left(2-\s[2]\right)}{(1-\e)^{5}\!\left(1-\s[2]\right)}\,\eln \\
&\RepQuad{9}\;\; - \pi\,\frac{\e^{2}(2-\e)^{2}\!\ssqrt}{(1-\e)^{5}\s\!\left(1-\s[2]\right)}\,\sln .
\end{split}
\end{align}
\end{subequations}
Plots of these two functions for several different values of \(\e = \{0, 0.1, 0.2, 0.5\}\) are provided in Fig.~\ref{fig:subluminalScalarCorr}.

For \(\e < 0\), the double integrals in eqs.~(\ref{eq:scalarIntegralEval}) could not be evaluated as the astrometric response is divergent at \(\Theta = \arcsec (1 - \e)\) (see Appendix~\ref{sec:AstrometricShift}). In order to find a meaningful interpretation of the correlation function \(\G_{ab}^{\mathsc{S}}\), one has to include the star terms in the astrometric response at each star, i.e. use eq.~(\ref{eq:astrometricShiftDef}) with eq.~(\ref{eq:astrometricShiftFull}). The resulting integrals would also depend on the distance to the star in units of \textsc{gw} wavelengths, \(d\):
\begin{subequations}
\begin{align}
\Gamma_{x\theta}^{\mathsc{S}} (\Theta, \e, d) = \int_{S^{2}} \!\!\!\dd\Omega_{\vect{q}}\;\delta n_{x}^{\mathsc{S}}(\vect{q}, d)\,\delta m_{\theta}^{\mathsc{S}}(\vect{q}, d) = \int_{0}^{2\pi} \!\!\!\!\dd\phi \int_{0}^{\pi} \!\!\!\dd\theta\,F_{x\theta}^{\mathsc{S}} (\Theta, \e, d, \theta, \phi) \,, \\
\Gamma_{y\phi}^{\mathsc{S}} (\Theta, \e, d) = \int_{S^{2}} \!\!\!\dd\Omega_{\vect{q}}\;\delta n_{y}^{\mathsc{S}}(\vect{q}, d)\,\delta m_{\phi}^{\mathsc{S}}(\vect{q}, d) = \int_{0}^{2\pi} \!\!\!\!\dd\phi \int_{0}^{\pi} \!\!\!\dd\theta\,F_{y\phi}^{\mathsc{S}} (\Theta, \e, d, \theta, \phi) \,.
\end{align}
\end{subequations}
These integrals, however, were not found to have analytical solutions and could only be evaluated numerically. Numerical integration (for \(\e = \{0, -0.1, -0.2, -0.5\}\) and \(d = 100\)) yields the curves shown in Fig.~\ref{fig:superluminalScalarCorr}.

\subsection{Vectorial correlations}
\label{sec:evalVectorialFunctions}
In this Appendix the evaluation of spatial correlation integrals for the vectorial \textsc{gw} polarization states, \(X\) and \(Y\), is briefly described. The integration is very similar to those for the \(+\) and \(\times\) states described in Appendix~\ref{sec:evalTensorialFunctions}.

Firstly, the \(x-\theta\) and \(y-\phi\) terms for the \(X\) polarized \textsc{gw} state are considered; the correlation integrals are defined in eqs.~(\ref{eq:GammaDef}), the vector \(\vect{q}\) is the direction on the sky from which the \textsc{gw} originates, and the components of the astrometric deflection may be evaluated in an analogous way to the one described in Appendix~\ref{sec:evalTensorialFunctions}. Therefore, these take the form
\begin{subequations}\label{eq:xIntegralEval}
\begin{align}
\Gamma_{x\theta}^{\mathsc{X}} (\Theta, \e) = \int_{S^{2}} \!\!\!\dd\Omega_{\vect{q}}\;\delta n_{x}^{\mathsc{X}}(\vect{q})\,\delta m_{\theta}^{\mathsc{X}}(\vect{q}) = \int_{0}^{2\pi} \!\!\!\!\dd\phi \int_{0}^{\pi} \!\!\!\dd\theta\,\sin\theta\,\delta n_{x}^{\mathsc{X}}(\e, \theta, \phi)\,\delta m_{\theta}^{\mathsc{X}}(\e, \theta, \phi) = \int_{0}^{2\pi} \!\!\!\!\dd\phi \int_{0}^{\pi} \!\!\!\dd\theta\,F_{x\theta}^{\mathsc{X}}(\Theta, \e, \theta, \phi) \,, \\
\Gamma_{y\phi}^{\mathsc{X}} (\Theta, \e) = \int_{S^{2}} \!\!\!\dd\Omega_{\vect{q}}\;\delta n_{y}^{\mathsc{X}}(\vect{q})\,\delta m_{\phi}^{\mathsc{X}}(\vect{q}) = \int_{0}^{2\pi} \!\!\!\!\dd\phi \int_{0}^{\pi} \!\!\!\dd\theta\,\sin\theta\,\delta n_{y}^{\mathsc{X}}(\e, \theta, \phi)\,\delta m_{\phi}^{\mathsc{X}}(\e, \theta, \phi) = \int_{0}^{2\pi} \!\!\!\!\dd\phi \int_{0}^{\pi} \!\!\!\dd\theta\,F_{y\phi}^{\mathsc{X}}(\Theta, \e, \theta, \phi) \,.
\end{align}
\end{subequations}
For \(\e > 0\), these two double integrals may now be evaluated by applying consecutively the techniques described in Appendices~\ref{app:azimuthalIntegral} and \ref{app:polarIntegral} to obtain functional forms for \(\Gamma^{\mathsc{X}}_{x\theta}(\Theta, \e)\) and \(\Gamma^{\mathsc{X}}_{y\phi}(\Theta, \e)\).

Secondly, the \(x-\theta\) and \(y-\phi\) terms for the \(Y\) polarized \textsc{gw} state are considered; the correlation integrals are defined in eqs.~(\ref{eq:GammaDef}), the vector \(\vect{q}\) is the direction on the sky from which the \textsc{gw} originates, and the components of the astrometric deflection may be evaluated in an analogous way to the one described in Appendix~\ref{sec:evalTensorialFunctions}. Therefore, these take the form
\begin{subequations}\label{eq:yIntegralEval}
\begin{align}
\Gamma_{x\theta}^{\mathsc{Y}} (\Theta, \e) = \int_{S^{2}} \!\!\!\dd\Omega_{\vect{q}}\;\delta n_{x}^{\mathsc{Y}}(\vect{q})\,\delta m_{\theta}^{\mathsc{Y}}(\vect{q}) = \int_{0}^{2\pi} \!\!\!\!\dd\phi \int_{0}^{\pi} \!\!\!\dd\theta\,\sin\theta\,\delta n_{x}^{\mathsc{Y}}(\e, \theta, \phi)\,\delta m_{\theta}^{\mathsc{Y}}(\e, \theta, \phi) = \int_{0}^{2\pi} \!\!\!\!\dd\phi \int_{0}^{\pi} \!\!\!\dd\theta\,F_{x\theta}^{\mathsc{Y}}(\Theta, \e, \theta, \phi) \,, \\
\Gamma_{y\phi}^{\mathsc{Y}} (\Theta, \e) = \int_{S^{2}} \!\!\!\dd\Omega_{\vect{q}}\;\delta n_{y}^{\mathsc{Y}}(\vect{q})\,\delta m_{\phi}^{\mathsc{Y}}(\vect{q}) = \int_{0}^{2\pi} \!\!\!\!\dd\phi \int_{0}^{\pi} \!\!\!\dd\theta\,\sin\theta\,\delta n_{y}^{\mathsc{Y}}(\e, \theta, \phi)\,\delta m_{\phi}^{\mathsc{Y}}(\e, \theta, \phi) = \int_{0}^{2\pi} \!\!\!\!\dd\phi \int_{0}^{\pi} \!\!\!\dd\theta\,F_{y\phi}^{\mathsc{Y}}(\Theta, \e, \theta, \phi) \,.
\end{align}
\end{subequations}
For \(\e > 0\), these two double integrals may now be evaluated by applying consecutively the techniques described in Appendices~\ref{app:azimuthalIntegral} and \ref{app:polarIntegral} to obtain functional forms for \(\Gamma^{\mathsc{Y}}_{x\theta} (\Theta, \e)\) and \(\Gamma^{\mathsc{Y}}_{y\phi}(\Theta, \e)\). For an unpolarised background containing equal power of both \(X\) and \(Y\) polarization states the combined spatial correlation functions may be defined as \(\Gamma^{\mathsc{X}, \mathsc{Y}}_{x\theta}(\Theta, \e) = \Gamma_{x\theta}^{\mathsc{X}} (\Theta, \e) + \Gamma_{x\theta}^{\mathsc{Y}} (\Theta, \e)\) and \(\Gamma^{\mathsc{X}, \mathsc{Y}}_{y\phi} (\Theta, \e) = \Gamma_{y\phi}^{\mathsc{X}} (\Theta, \e) + \Gamma_{y\phi}^{\mathsc{Y}} (\Theta, \e)\). These new functions may be evaluated by taking the sums of the expression derived above to give the results
\begin{subequations}
\begin{align}
\begin{split}\label{eq:SublimunalVectorialxthCorr}
&\G_{x\theta}^{\mathsc{X}, \mathsc{Y}} (\Theta, \e) = \frac{2\pi}{3}\,\frac{\left(2+18\e-5\e^{2}-4\e^{3}+\e^{4}\right) + 4\!\left(1-8\e+4\e^{2}\right)\!\s[2]}{(1-\e)^{4}} \\
&\RepQuad{5}- \pi\,\frac{\e\!\left(6+\e-4\e^{2}+\e^{3}\right) + \left(4-18\e+5\e^{2}+4\e^{3}-\e^{4}\right)\!\s[2] + 4\e(2-\e)\s[4]}{(1-\e)^{5}\!\left(1-\s[2]\right)}\,\eln \\
&\RepQuad{5}+ 4\pi\,\frac{\left(\e(2-\e) + 2(1-\e)^{2}\s[2]\right)\!\s}{(1-\e)^{5}\!\left(1-\s[2]\right)\!\ssqrt}\,\sln , 
\end{split}\\[10pt]
\begin{split}\label{eq:SublimunalVectorialyphCorr}
&\G_{y\phi}^{\mathsc{X}, \mathsc{Y}} (\Theta, \e) = \frac{2\pi}{3}\,\frac{\left(2+6\e+\e^{2}-4\e^{3}+\e^{4}\right) + 2(1-\e)^{2}\!\left(2+2\e-\e^{2}\right)\!\s[2]}{(1-\e)^{4}} \\
&\RepQuad{5}- \pi\,\frac{\e(1+\e)(3-\e)(2-\e) + \left(4-6\e-9\e^{2}+12\e^{3}-3\e^{4}\right)\!\s[2] - 2\e(2-\e)(1-\e)^{2}\s[4]}{(1-\e)^{5}\!\left(1-\s[2]\right)}\,\eln \\
&\RepQuad{5}+ 4\pi\,\frac{\left(\e(2-\e) + 2(1-\e)^{2}\s[2]\right)\!\ssqrt}{(1-\e)^{5}\s\!\left(1-\s[2]\right)}\times \\
&\RepQuad{27} \times\sln . 
\end{split}
\end{align}
\end{subequations}
Plots of these two functions for several different values of \(\e = \{0, 0.1, 0.2, 0.5\}\) are provided in Fig.~\ref{fig:subluminalVectorialCorr}.

For \(\e < 0\), the double integrals in eqs.~(\ref{eq:xIntegralEval}) and (\ref{eq:yIntegralEval}) could not be evaluated as the astrometric response is divergent at \(\Theta = \arcsec (1 - \e)\) (see Appendix~\ref{sec:AstrometricShift}). In order to find a meaningful interpretation of the correlation function \(\G_{ab}^{\mathsc{X}, \mathsc{Y}}\), one has to include the star terms in the astrometric response at each star, i.e. use eq.~(\ref{eq:astrometricShiftDef}) with eq.~(\ref{eq:astrometricShiftFull}). The resulting integrals would also depend on the distance to the star in units of \textsc{gw} wavelengths, \(d\):
\begin{subequations}\label{eq:xyIntegralEval}
\begin{align}
\Gamma_{x\theta}^{\mathsc{X}, \mathsc{Y}} (\Theta, \e, d) = \int_{S^{2}} \!\!\!\dd\Omega_{\vect{q}}\;\delta n_{x}^{\mathsc{X}, \mathsc{Y}}(\vect{q}, d)\,\delta m_{\theta}^{\mathsc{X}, \mathsc{Y}}(\vect{q}, d) = \int_{0}^{2\pi} \!\!\!\!\dd\phi \int_{0}^{\pi} \!\!\!\dd\theta\,F_{x\theta}^{\mathsc{X}, \mathsc{Y}}(\Theta, \e, d, \theta, \phi) \,, \\
\Gamma_{y\phi}^{\mathsc{X}, \mathsc{Y}} (\Theta, \e, d) = \int_{S^{2}} \!\!\!\dd\Omega_{\vect{q}}\;\delta n_{y}^{\mathsc{X}, \mathsc{Y}}(\vect{q}, d)\,\delta m_{\phi}^{\mathsc{X}, \mathsc{Y}}(\vect{q}, d) = \int_{0}^{2\pi} \!\!\!\!\dd\phi \int_{0}^{\pi} \!\!\!\dd\theta\,F_{y\phi}^{\mathsc{X}, \mathsc{Y}}(\Theta, \e, d, \theta, \phi) \,.
\end{align}
\end{subequations}
These integrals, however, were not found to have analytical solutions and could only be evaluated numerically. Numerical integration (for \(\e = \{0, -0.1, -0.2, -0.5\}\) and \(d = 100\)) yields the curves shown in Fig.~\ref{fig:superluminalVectorialCorr}.

\subsection{Scalar longitudinal correlations}
\label{sec:evalLongitudinalFunctions}
\noindent
In this Appendix the evaluation of spatial correlation integrals for the transverse scalar \textsc{gw} polarization state, \(L\), is briefly described. The \(x-\theta\) and \(y-\phi\) correlation integrals are defined in eqs.~(\ref{eq:GammaDef}), the vector \(\vect{q}\) is the direction on the sky from which the \textsc{gw} originates, and the components of the astrometric deflection may be evaluated in an analogous way to the one described in Appendix~\ref{sec:evalTensorialFunctions}. Therefore, these take the form
\begin{subequations}\label{eq:longitudinalIntegralEval}
\begin{align}
\Gamma_{x\theta}^{\mathsc{L}} (\Theta, \e) = \int_{S^{2}} \!\!\!\dd\Omega_{\vect{q}}\;\delta n_{x}^{\mathsc{L}}(\vect{q})\,\delta m_{\theta}^{\mathsc{L}}(\vect{q}) = \int_{0}^{2\pi} \!\!\!\!\dd\phi \int_{0}^{\pi} \!\!\!\dd\theta\,\sin\theta\,\delta n_{x}^{\mathsc{L}}(\e, \theta, \phi)\,\delta m_{\theta}^{\mathsc{L}}(\e, \theta, \phi) = \int_{0}^{2\pi} \!\!\!\!\dd\phi \int_{0}^{\pi} \!\!\!\dd\theta\,F_{x\theta}^{\mathsc{L}}(\Theta, \e, \theta, \phi) \,,  \label{eq:SublimunalLongitudinalxthCorr} \\
\Gamma_{y\phi}^{\mathsc{L}} (\Theta, \e) = \int_{S^{2}} \!\!\!\dd\Omega_{\vect{q}}\;\delta n_{y}^{\mathsc{L}}(\vect{q})\,\delta m_{\phi}^{\mathsc{L}}(\vect{q}) = \int_{0}^{2\pi} \!\!\!\!\dd\phi \int_{0}^{\pi} \!\!\!\dd\theta\,\sin\theta\,\delta n_{y}^{\mathsc{L}}(\e, \theta, \phi)\,\delta m_{\phi}^{\mathsc{L}}(\e, \theta, \phi) = \int_{0}^{2\pi} \!\!\!\!\dd\phi \int_{0}^{\pi} \!\!\!\dd\theta\,F_{y\phi}^{\mathsc{L}}(\Theta, \e, \theta, \phi) \,. \label{eq:SublimunalLongitudinalyphCorr}
\end{align}
\end{subequations}
For \(\e > 0\), these two double integrals may now be evaluated by applying consecutively the techniques described in Appendices~\ref{app:azimuthalIntegral} and \ref{app:polarIntegral} to obtain functional forms for \(\Gamma^{\mathsc{L}}_{x\theta}(\Theta, \e)\) and \(\Gamma^{\mathsc{L}}_{y\phi}(\Theta, \e)\):
\begin{subequations}
\begin{align}
\begin{split}
&\G_{x\theta}^{\mathsc{L}} (\Theta, \e) = -\frac{2\pi}{3}\,\frac{\left(5+2\e-\e^{2}\right) - 2\!\left(2+2\e-\e^{2}\right)\!\s[2]}{(1-\e)^{4}} \\
&\RepQuad{5}+ \pi\,\frac{\left(1+2\e-\e^{2}\right) -3\e(2-\e)\s[2] + 2\e(2-\e)\s[4]}{(1-\e)^{5}\!\left(1-\s[2]\right)}\,\eln \\
&\RepQuad{5}- 2\pi\,\frac{\s}{(1-\e)^{5}\!\left(1-\s[2]\right)\!\ssqrt}\,\sln ,
\end{split}\\[10pt]
\begin{split}
&\G_{y\phi}^{\mathsc{L}} (\Theta, \e) = -\frac{2\pi}{3}\,\frac{2+2\e-\e^{2}}{(1-\e)^{4}} + \pi\,\frac{\left(1+2\e-\e^{2}\right) - \e(2-\e)\s[2]}{(1-\e)^{5}\!\left(1-\s[2]\right)}\,\eln \\
&\RepQuad{13}\;\; - 2\pi\,\frac{\ssqrt}{(1-\e)^{5}\s\!\left(1-\s[2]\right)}\,\sln .
\end{split}
\end{align}
\end{subequations}
Plots of these two functions for several different values of \(\e = \{0, 0.1, 0.2, 0.5\}\) are provided in Fig.~\ref{fig:subluminalLongitudinalCorr}.

For \(\e < 0\), the double integrals in eqs.~(\ref{eq:longitudinalIntegralEval}) could not be evaluated as the astrometric response is divergent at \(\Theta = \arcsec (1 - \e)\) (see Appendix~\ref{sec:AstrometricShift}). In order to find a meaningful interpretation of the correlation function \(\G_{ab}^{\mathsc{L}}\), one has to include the star terms in the astrometric response at each star, i.e. use eq.~(\ref{eq:astrometricShiftDef}) with eq.~(\ref{eq:astrometricShiftFull}). The resulting integrals would also depend on the distance to the star in units of \textsc{gw} wavelengths, \(d\):
\begin{subequations}
\begin{align}
\Gamma_{x\theta}^{\mathsc{L}} (\Theta, \e, d) = \int_{S^{2}} \!\!\!\dd\Omega_{\vect{q}}\;\delta n_{x}^{\mathsc{L}}(\vect{q}, d)\,\delta m_{\theta}^{\mathsc{L}}(\vect{q}, d) = \int_{0}^{2\pi} \!\!\!\!\dd\phi \int_{0}^{\pi} \!\!\!\dd\theta\,F_{x\theta}^{\mathsc{L}} (\Theta, \e, d, \theta, \phi) \,, \\
\Gamma_{y\phi}^{\mathsc{L}} (\Theta, \e, d) = \int_{S^{2}} \!\!\!\dd\Omega_{\vect{q}}\;\delta n_{y}^{\mathsc{L}}(\vect{q}, d)\,\delta m_{\phi}^{\mathsc{L}}(\vect{q}, d) = \int_{0}^{2\pi} \!\!\!\!\dd\phi \int_{0}^{\pi} \!\!\!\dd\theta\,F_{y\phi}^{\mathsc{L}} (\Theta, \e, d, \theta, \phi) \,.
\end{align}
\end{subequations}
These integrals, however, were not found to have analytical solutions and could only be evaluated numerically. Numerical integration (for \(\e = \{0, -0.1, -0.2, -0.5\}\) and \(d = 100\)) yields the curves shown in Fig.~\ref{fig:superluminalLongitudinalCorr}.

\section{Angular Power Spectra for Luminal GWs}
\label{app:luminalSpectra}
\noindent
The angular power spectrum coefficients can be derived analytically through eqs~(\ref{eq:spectraCoeffsFormulae}) (as a function of the mode number \(\ell\)) for the case when the gravitational wave is propagating at the speed of light. Closed-form expressions cannot be derived in the case when \(\e > 0\), although analytical solutions can still be found for specific \(\ell\). For \(\e < 0\), analytical solutions are not possible, as the correlation functions are tabulated numerically.

\subsection{Tensorial transverse-traceless power spectra}
\label{app:tensorialLuminalSpectra}
\noindent
In the luminal (\(\e=0\)) case, the transverse traceless correlation functions \(\Gamma_{x\theta}^{+, \times} (\Theta) = \Gamma_{y\phi}^{+, \times} (\Theta)\) are equal (cf. eq.~(45) from \citep{2018PhRvD..97l4058M}), therefore the respective \(E\) and \(B\) coefficients would be equal, and in turn equal to \(C_{\ell}^{+, \times}\). Using this, eqs.~(\ref{eq:spectraCoeffsFormulae}) become
\begin{align}
C_{\ell}^{+, \times} = C_{\ell}^{+, \times, \mathsc{E}} = C_{\ell}^{+, \times, \mathsc{B}} = \frac{1}{2\ell (\ell+1)} \bigintsss_{0}^{\pi} \!\!\!\!\dd\Theta \, \sin\Theta \, P_{\ell}\!\left(\cos\Theta\right)\!\left(\!\frac{16 \pi}{3} - \frac{8 \pi}{3}\,\s[2] + 32\pi\,\s[2] \lns\!\right).
\end{align}
The \(\Theta\) integral could be solved exactly, as a function of the multipole order \(\ell\) to give
\begin{align}
C_{\ell}^{+, \times} = \frac{\pi}{3}\,\frac{2^{\ell+1}}{\ell(\ell + 1)}\,\sum_{k = 0}^{\ell} \binom{\ell}{k} \binom{\sfrac{1}{2}\left(\ell + k - 1\right)}{\ell} \frac{(-1)^{\ell} (2k - 1) + 13 + 4 k - 12 (k + 2) \, \pFq{3}{2}{\!(1,1,-k-1)}{(2,2)}{2}}{(k+1) (k+2)}\,,
\end{align}
where \(_{p}F_{q}\) is the generalized hypergeometric function. The first 20 coefficients of this sequence are plotted in Fig.~\ref{fig:subluminalTensorialCoeffs} and are tabulated in Table~\ref{tab:coefficients}.

\subsection{Scalar transverse power spectra}
\label{app:scalarLuminalSpectra}
\noindent
In the luminal (\(\e=0\)) case, the transverse scalar correlation functions \(\Gamma_{x\theta}^{\mathsc{S}} (\Theta)\) and \(\Gamma_{y\phi}^{\mathsc{S}} (\Theta)\) are given by eqs. (47) of \citep{2018PhRvD..97l4058M}. Using eqs.~(\ref{eq:spectraCoeffsFormulae}), since \(\Gamma_{y\phi}^{+, \times} (\Theta)\) is in fact a constant (independent of \(\Theta\)), the \(B\) coefficients vanish identically for all \(\ell\), and the \(C_{\ell}^{\mathsc{S}}\) are given by
\begin{align}
C_{\ell}^{\mathsc{S}} = \frac{C_{\ell}^{\mathsc{S}, \mathsc{E}}}{\sqrt{2}} = \frac{1}{2\ell (\ell+1)} \bigintsss_{0}^{\pi} \!\!\!\!\dd\Theta \, \sin\Theta \, P_{\ell}\!\left(\cos\Theta\right)\!\left(\!\frac{4 \pi}{3}\!\left(1 - 2 \s[2]\right)\!\right).
\end{align}
The \(\Theta\) integral is straightforward to evaluate in terms of the multipole number \(\ell\), yielding
\begin{align}
C_{\ell}^{\mathsc{S}} = \frac{\pi}{3}\,\frac{2^{\ell+1/2}}{\ell(\ell + 1)}\,\sum_{k = 0}^{\ell} \binom{\ell}{k} \binom{\sfrac{1}{2}\left(\ell + k - 1\right)}{\ell} \frac{1 + (-1)^{k+1}}{k+2}\,.
\end{align}
This expression is non-zero only for \(\ell = 1\), corresponding to a dipole in the astrometric pattern on the sky. This spectrum is plotted in Fig.~\ref{fig:subluminalScalarCoeffs} and tabulated in Table~\ref{tab:coefficients}.

\subsection{Vectorial power spectra}
\label{app:vectorialLuminalSpectra}
\noindent
In the luminal (\(\e=0\)) case, the vectorial correlation functions \(\Gamma_{x\theta}^{\mathsc{X}, \mathsc{Y}} (\Theta) = \Gamma_{y\phi}^{\mathsc{X}, \mathsc{Y}} (\Theta)\) are equal (cf. eq.~(49) from \citep{2018PhRvD..97l4058M}), therefore the respective \(E\) and \(B\) coefficients would be equal, and in turn equal to \(C_{\ell}^{\mathsc{X}, \mathsc{Y}}\). Using this, eqs.~(\ref{eq:spectraCoeffsFormulae}) become
\begin{align}
C_{\ell}^{\mathsc{X}, \mathsc{Y}} = C_{\ell}^{\mathsc{X}, \mathsc{Y}, \mathsc{E}} = C_{\ell}^{\mathsc{X}, \mathsc{Y}, \mathsc{B}} = \frac{1}{2\ell (\ell+1)} \bigintsss_{0}^{\pi} \!\!\!\!\dd\Theta \, \sin\Theta \, P_{\ell}\!\left(\cos\Theta\right)\!\left(\!-\frac{28 \pi}{3} + \frac{32 \pi}{3}\,\s[2] - 8\pi \lns\!\right).
\end{align}
The \(\Theta\) integral could be solved exactly, as a function of the multipole order \(\ell\) to give
\begin{align}
C_{\ell}^{\mathsc{X}, \mathsc{Y}} = \frac{\pi}{3}\,\frac{2^{\ell+1}}{\ell (\ell + 1)}\,\sum_{k = 0}^{\ell} \binom{\ell}{k} \binom{\sfrac{1}{2}\left(\ell + k - 1\right)}{\ell} \frac{(-1)^{k} (k - 2) - 7k - 10 + 3 \left(2 H_{2\lfloor k/2 \rfloor + 1} - H_{\lfloor k/2 \rfloor }\right)}{(k+1) (k+2)}\,,
\end{align}
where \(H_{n}\) is the \(n\)-th harmonic number, and \(\lfloor x \rfloor\) is the floor function. The first 20 coefficients of this sequence are plotted in Fig.~\ref{fig:subluminalVectorialCoeffs} and are tabulated in Table~\ref{tab:coefficients}.

\subsection{Scalar longitudinal power spectra}
\label{app:longitudinalLuminalSpectra}
\noindent
Naively plugging the (unphysical) \(\Gamma_{x\theta}^{\mathsc{L}} (\Theta)\) and \(\Gamma_{y\phi}^{\mathsc{L}} (\Theta)\) (given by eq.~(51) of \citep{2018PhRvD..97l4058M}) into eqs.~(\ref{eq:spectraCoeffsFormulae}) yields negative \(C_{\ell}^{\mathsc{L}, \mathsc{E}}\) power spectrum coefficients (and vanishing \(C_{\ell}^{\mathsc{L}, \mathsc{B}}\)). These results are clearly unphysical as variance cannot be negative. An alternative approach, involving the resolution of the astrometric response along \textsc{vsh} before calculating the correlation coefficients between them produces reliable and physical results. The derivation of these results is beyond the scope of this article, however the first 20 coefficients of this sequence are plotted in Fig.~\ref{fig:subluminalLongitudinalCoeffs} and are tabulated in Table~\ref{tab:coefficients}.

\begin{table}[t]
\begin{tabular}{rcccccccccc}
\toprule
\(\ell\) & 1 & 2 & 3 & 4 & 5 & 6 & 7 & 8 & 9 & 10 \\
\midrule
\(C_{\ell}^{+, \times} / \pi\) & 0 & 1/9 & 1/90 & 1/450 & 1/1575 & 1/4410 & 1/10584 & 1/22680 & 1/44550 & 1/81675 \\[2pt]
\(C_{\ell}^{\mathsc{S}, \mathsc{E}} / \pi\) & 2/9 & 0 & 0 & 0 & 0 & 0 & 0 & 0 & 0 & 0 \\[2pt]
\(C_{\ell}^{\mathsc{X}, \mathsc{Y}} / \pi\) & 1/9 & 1/9 & 1/36 & 1/100 & 1/225 & 1/441 & 1/784 & 1/1296 & 1/2025 & 1/3025 \\[2pt]
\(C_{\ell}^{\mathsc{L}, \mathsc{E}} / \pi\) & 1/9 & 1/3 & 1/6 & 1/10 & 1/15 & 1/21 & 1/28 & 1/36 & 1/45 & 1/55 \\[2pt]
\toprule
\(\ell\) & 11 & 12 & 13 & 14 & 15 & 16 & 17 & 18 & 19 & 20 \\
\midrule
\(C_{\ell}^{+, \times} / \pi\) & 1/141570 & 1/234234 & 1/372645 & 1/573300 & 1/856800 & 1/1248480 & 1/1779084 & 1/2485485 & 1/3411450 & 1/4608450 \\[2pt]
\(C_{\ell}^{\mathsc{S}, \mathsc{E}} / \pi\) & 0 & 0 & 0 & 0 & 0 & 0 & 0 & 0 & 0 & 0 \\[2pt]
\(C_{\ell}^{\mathsc{X}, \mathsc{Y}} / \pi\) & 1/4356 & 1/6084 & 1/8281 & 1/11025 & 1/14400 & 1/18496 & 1/23409 & 1/29241 & 1/36100 & 1/44100 \\[2pt]
\(C_{\ell}^{\mathsc{L}, \mathsc{E}} / \pi\) & 1/66 & 1/78 & 1/91 & 1/105 & 1/120 & 1/136 & 1/153 & 1/171 & 1/190 & 1/210 \\
\bottomrule
\end{tabular}
\caption{The first 20 angular power spectrum coefficients for each polarization content in the case \(\e = 0\). The closed form expressions for \(C_{\ell}^{+, \times}\), \(C_{\ell}^{\mathsc{S}}\), and \(C_{\ell}^{\mathsc{X}, \mathsc{Y}}\) are given in Appendices~\ref{app:tensorialLuminalSpectra}, \ref{app:scalarLuminalSpectra}, and \ref{app:vectorialLuminalSpectra}, respectively. The transverse scalar power spectrum has a single non-zero coefficient, \(\ell = 2\), corresponding to a dipole on the sky. The longitudinal scalar power spectrum was derived through another method, which is described briefly in Appendix~\ref{app:longitudinalLuminalSpectra}.}
\label{tab:coefficients}
\end{table}

\begin{figure}[b]
\includegraphics[scale=1]{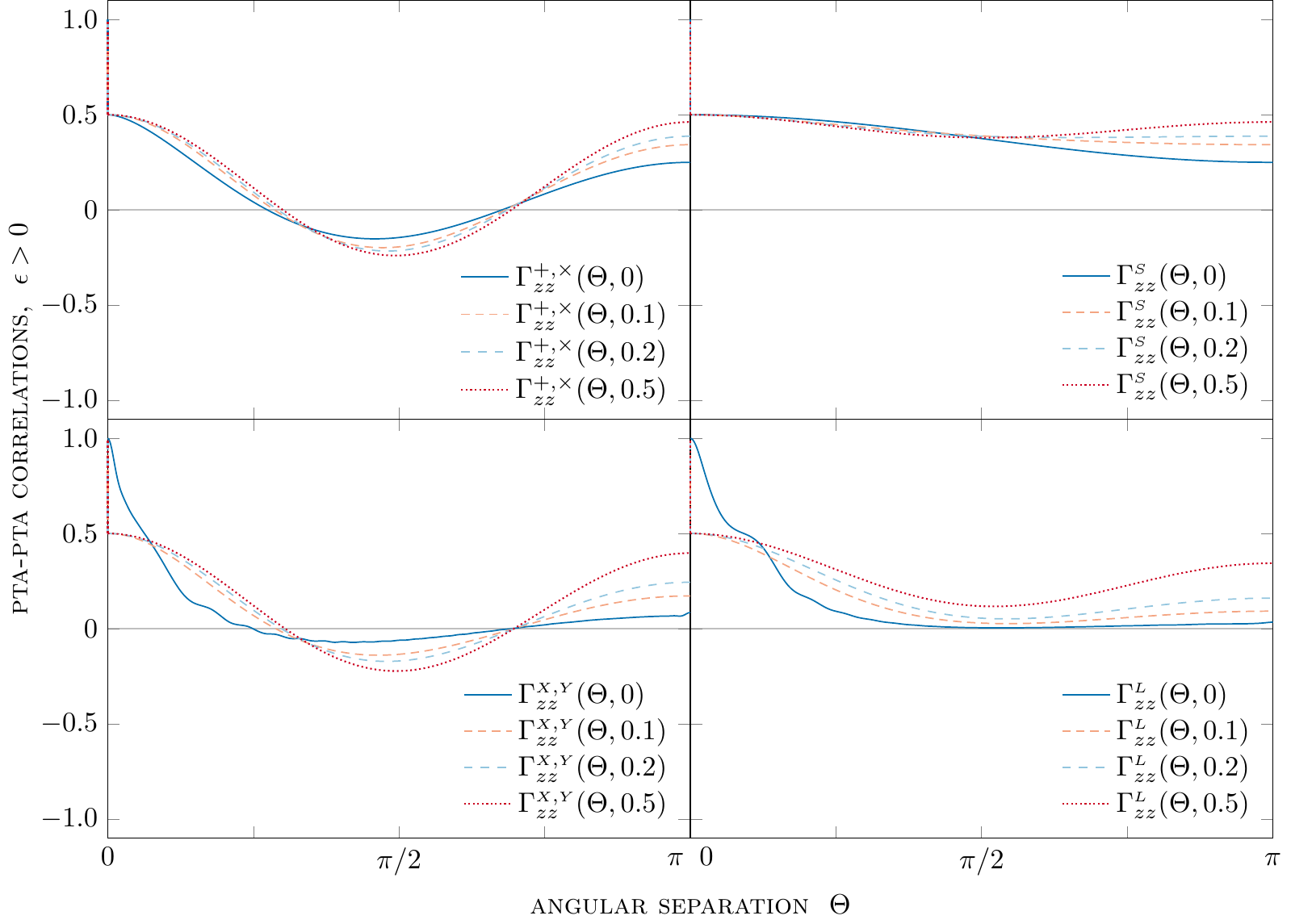}
\caption{The sub-luminal \textsc{pta}-\textsc{pta} correlations for 4 different values of \(\e \in \{0, 0.1, 0.2, 0.5\}\). \emph{Upper left panel:} The tensorial transverse-traceless correlation function \(\G_{zz}^{+, \times} (\Theta, \e) = \G_{zz}^{+} (\Theta, \e)\) given by eq.~(\ref{eq:ptaPtaTensorial}). The \(\G_{zz}^{\times} (\Theta, \e)\) correlation vanishes. The curve \(\G_{zz}^{+, \times} (\Theta, 0)\) is the well-known Hellings-Downs curve \citep{1983ApJ...265L..39H}. \emph{Upper right panel:} The scalar transverse correlation function \(\G_{zz}^{\mathsc{S}} (\Theta, \e)\) given by eq.~(\ref{eq:ptaPtaScalar}). \emph{Lower left panel:} The vectorial correlation function \(\G_{zz}^{\mathsc{X}, \mathsc{Y}} (\Theta, \e) = \G_{zz}^{\mathsc{X}} (\Theta, \e)\) given by eq.~(\ref{eq:ptaPtaVectorial}). The \(\G_{zz}^{\mathsc{Y}} (\Theta, \e)\) correlation vanishes. The correlation \(\G_{zz}^{\mathsc{X}, \mathsc{Y}} (\Theta, 0)\) is computed numerically with both pulsars placed at \(d = 100\). \emph{Lower right panel:} The scalar longitudinal correlation function \(\G_{zz}^{\mathsc{L}} (\Theta, \e)\) given by eq.~(\ref{eq:ptaPtaLongitudinal}). The correlation \(\G_{zz}^{\mathsc{L}} (\Theta, 0)\) is computed numerically with both pulsars placed at \(d = 100\).}
\label{fig:redshiftCorrelation}
\end{figure}

\section{PTA-PTA correlations}
\label{sec:AppRedshift}
\noindent
The \textsc{pta}-\textsc{pta} correlation functions in the sub-luminal regime, defined by eq.~(\ref{eq:GammazzDef}) in Section~\ref{sec:redshiftResults}, are presented in analytical form for the first time in this appendix. The integrals were performed in an analogous fashion to the method presented in Appendix~\ref{app:evalFunctions}, and are therefore not described in detail here.

\subsection{Tensorial transverse-traceless correlations}
\noindent
The \textsc{pta}-\textsc{pta} correlation for a background of sub-luminal tensorial transverse-traceless \textsc{gw}s \(\G_{zz}^{+, \times} (\Theta, \e)\) is given below. The correlation function \(\G_{zz}^{\times} (\Theta, \e)\) vanishes identically, so the correlation is given solely by the function \(\G_{zz}^{+} (\Theta, \e)\). The luminal limit (\(\e = 0\)) of this function is the Hellings-Downs curve \citep{1983ApJ...265L..39H}.
\begin{align}
\begin{split}\label{eq:ptaPtaTensorial}
&\G_{zz}^{+, \times} (\Theta, \e) = \frac{\pi}{3}\,\frac{\left(4 + 10\e - 5\e^{2}\right) - 2\left(1 + 10\e - 5\e^{2}\right)\!\s[2]}{(1-\e)^{4}} - 2\pi\,\frac{\e(2-\e) + \left(2 - 6\e + 3\e^{2}\right)\!\s[2]}{(1-\e)^{5}}\,\eln \\
&\RepQuad{3} + \pi\,\frac{\e^{2}(2-\e)^{2} + 8\e(2-\e)(1-\e)^{2}\s[2] + 8(1 - \e)^{4}\s[4]}{(1-\e)^{5}\s\ssqrt}\,\sln
\end{split}
\end{align}
This correlation function is plotted in the upper left panel of Fig.~\ref{fig:redshiftCorrelation}.

\subsection{Scalar transverse correlations}
\noindent
The \textsc{pta}-\textsc{pta} correlation for a background of sub-luminal scalar transverse \textsc{gw}s \(\G_{zz}^{\mathsc{S}} (\Theta, \e)\) is given below.
\begin{align}
\begin{split}\label{eq:ptaPtaScalar}
&\G_{zz}^{\mathsc{S}} (\Theta, \e) = \frac{\pi}{3}\,\frac{\left(4 + 10\e - 5\e^{2}\right) - 2\left(1 + 10\e - 5\e^{2}\right)\!\s[2]}{(1-\e)^{4}} - 2\pi\,\frac{\e(2 - \e)\!\left(1 - \s[2]\right)}{(1-\e)^{5}}\,\eln \\
&\RepQuad{8}\;\;\;\; + \pi\,\frac{\e^{2}(2-\e)^{2}}{(1-\e)^{5}\s\ssqrt}\,\sln .
\end{split}
\end{align}
This correlation function is plotted in the upper right panel of Fig.~\ref{fig:redshiftCorrelation}.

\begin{figure}[b]
\includegraphics[scale=1]{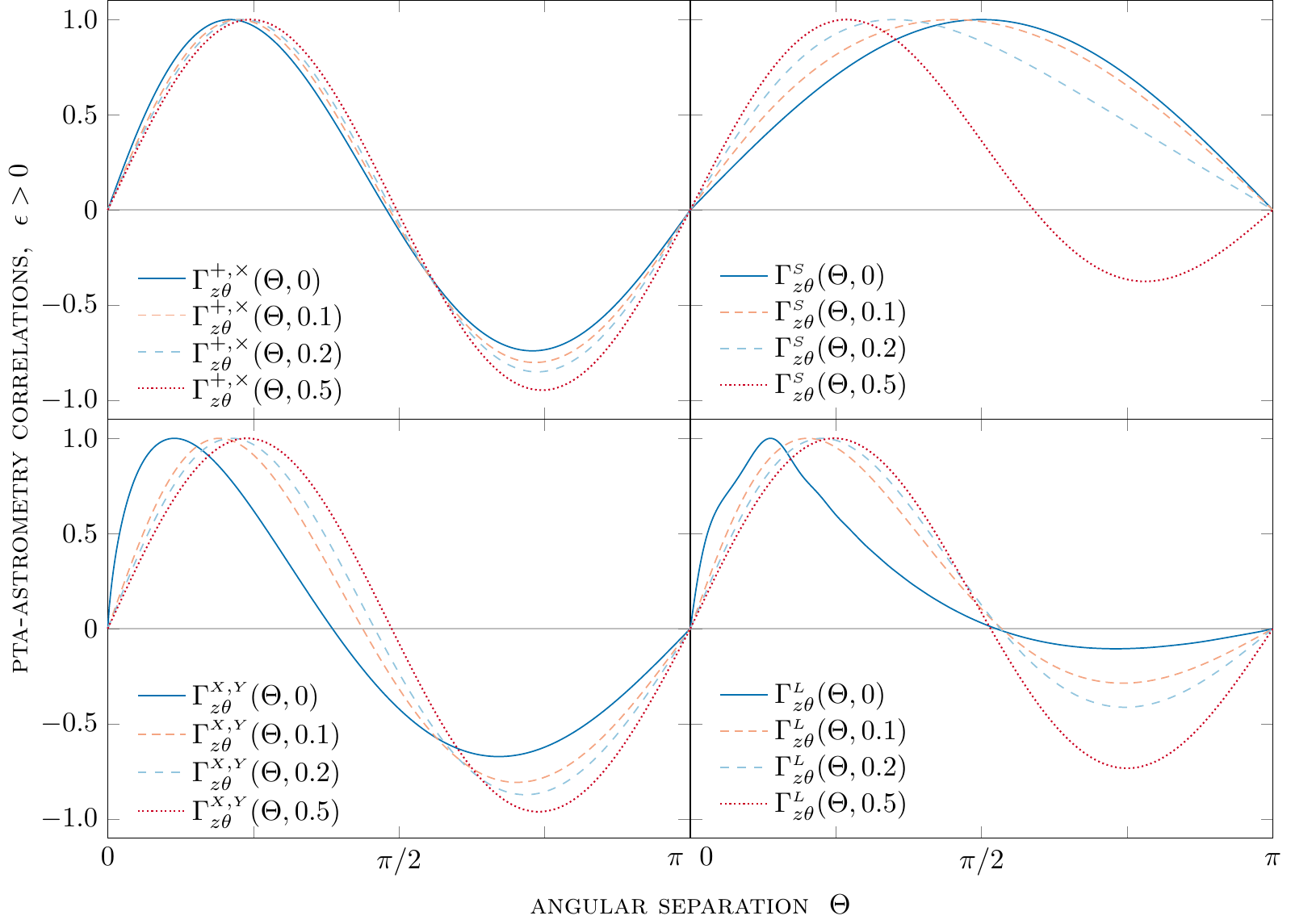}
\caption{The sub-luminal \textsc{pta}-astrometry correlations cases for 4 different values of \(\e \in \{0, 0.1, 0.2, 0.5\}\). The \(\G_{z\phi}^{\mathsc{P}} (\Theta, \e)\) correlations vanish for all polarization contents \(P\). \emph{Upper left panel:} The tensorial transverse-traceless correlation function \(\G_{z\theta}^{+, \times} (\Theta, \e) = \G_{z\theta}^{+} (\Theta, \e)\) given by eq.~(\ref{eq:ptaAstrometryTensorial}). The \(\G_{z\theta}^{\times} (\Theta, \e)\) correlation vanishes. \emph{Upper right panel:} The scalar transverse correlation function \(\G_{z\theta}^{\mathsc{S}} (\Theta, \e)\) given by eq.~(\ref{eq:ptaAstrometryScalar}). \emph{Lower left panel:} The vectorial correlation function \(\G_{z\theta}^{\mathsc{X}, \mathsc{Y}} (\Theta, \e) = \G_{z\theta}^{\mathsc{X}} (\Theta, \e)\) given by eq.~(\ref{eq:ptaAstrometryVectorial}). The \(\G_{z \theta}^{\mathsc{Y}} (\Theta, \e)\) correlation vanishes. \emph{Lower right panel:} The scalar longitudinal correlation function \(\G_{z\theta}^{\mathsc{S}} (\Theta, \e)\) given by eq.~(\ref{eq:ptaAstrometryLongitudinal}).}
\label{fig:redshiftAstrometryCorrelation}
\end{figure}

\subsection{Vectorial correlations}
\noindent
The \textsc{pta}-\textsc{pta} correlation for a background of sub-luminal vectorial \textsc{gw}s \(\G_{zz}^{\mathsc{X}, \mathsc{Y}} (\Theta, \e)\) is given below. The correlation function \(\G_{zz}^{\mathsc{Y}} (\Theta, \e)\) vanishes identically, so the correlation is given solely by the function \(\G_{zz}^{\mathsc{X}} (\Theta, \e)\). The luminal limit (\(\e = 0\)) of this function does not exist unless the pulsar terms are included, and the correlation cannot be derived analytically.
\begin{align}
\begin{split}\label{eq:ptaPtaVectorial}
&\G_{zz}^{\mathsc{X}, \mathsc{Y}} (\Theta, \e) = -\frac{4\pi}{3}\,\frac{\left(7 + 4\e - 2\e^{2}\right) - 4\left(2 + 2\e - \e^{2}\right)\!\s[2]}{(1-\e)^{4}} + 4\pi\,\frac{\left(1 + 2\e - \e^{2}\right) - 2\e(2 - \e)\s[2]}{(1-\e)^{5}}\,\eln \\
&\RepQuad{8}\;\; - 4\pi\,\frac{\e(2-\e) + 2 (1 - \e)^{2}\s[2]}{(1-\e)^{5}\s\ssqrt}\,\sln
\end{split}
\end{align}
This correlation function is plotted in the lower left panel of Fig.~\ref{fig:redshiftCorrelation}.

\subsection{Scalar longitudinal correlations}
\noindent
The \textsc{pta}-\textsc{pta} correlation for a background of sub-luminal scalar longitudinal \textsc{gw}s \(\G_{zz}^{\mathsc{L}} (\Theta, \e)\) is given below. The luminal limit (\(\e = 0\)) of this function does not exist unless the pulsar terms are included, and the correlation cannot be derived analytically.
\begin{align}
\begin{split}\label{eq:ptaPtaLongitudinal}
&\G_{zz}^{\mathsc{L}} (\Theta, \e) = \frac{2\pi}{3}\,\frac{\left(10 - 2\e + \e^{2}\right) - 2\left(7 - 2\e + \e^{2}\right)\!\s[2]}{(1-\e)^{4}} - 4\pi\,\frac{1-\s[2]}{(1-\e)^{5}}\,\eln \\
&\RepQuad{8} + 2\pi\,\frac{1}{(1-\e)^{5}\s\ssqrt}\,\sln
\end{split}
\end{align}
This correlation function is plotted in the lower right panel of Fig.~\ref{fig:redshiftCorrelation}.

\section{PTA-astrometry Correlations}
\label{sec:AppRedshiftAstrometry}
\noindent
The \textsc{pta}-astrometry correlation functions were proposed and examined in the luminal case in \citep{2018PhRvD..97l4058M}. These results were extended to the sub-luminal case, as defined by eq.~(\ref{eq:GammazbDef}) in Section~\ref{sec:redshiftAstrometryResults}. The integrals were performed in an analogous fashion to the method presented in Appendix~\ref{app:evalFunctions}, and are therefore not described in detail here. All correlations of the form \(\G_{z\phi}^{\mathsc{P}} (\Theta, \e)\) vanish.

\subsection{Tensorial transverse-traceless correlations}
\noindent
The \textsc{pta}-astrometry correlation for a background of sub-luminal tensorial transverse-traceless \textsc{gw}s \(\G_{z\theta}^{+, \times} (\Theta, \e)\) is given below. The correlation function \(\G_{z\theta}^{\times} (\Theta, \e)\) vanishes identically, so the correlation is given solely by the function \(\G_{z\theta}^{+} (\Theta, \e)\).
\begin{align}
\begin{split}\label{eq:ptaAstrometryTensorial}
&\G_{z\theta}^{+, \times} (\Theta, \e) = \frac{\pi}{3}\,\frac{\left(8-10\e+5\e^{2}\right)\!\s\sqrt{1-\s[2]}}{(1-\e)^{4}} \\
&\RepQuad{5} - \frac{\pi}{2}\,\frac{\left(2\e\left(4-6\e+4\e^{2}-\e^{3}\right) + \left(8-24\e+24\e^{2}-12\e^{3}+3\e^{4}\right)\!\s[2]\right)\!\s}{(1-\e)^{5}\!\sqrt{1-\s[2]}}\,\eln \\
&\RepQuad{5} + \pi\,\frac{\e^{2}(2-\e)^{2} + 8\e(2-\e)(1-\e)^{2}\s[2] + 8(1-\e)^{4}\s[4]}{(1-\e)^{5}\!\sqrt{1-\s[2]} \ssqrt}\,\times \\
&\RepQuad{27} \times\sln
\end{split}
\end{align}
This correlation function is plotted in the upper left panel of Fig.~\ref{fig:redshiftAstrometryCorrelation}.

\subsection{Scalar transverse correlations}
\noindent
The \textsc{pta}-astrometry correlation for a background of sub-luminal scalar transverse \textsc{gw}s \(\G_{z\theta}^{\mathsc{S}} (\Theta, \e)\) is given below.
\begin{align}
\begin{split}\label{eq:ptaAstrometryScalar}
&\G_{z\theta}^{\mathsc{S}} (\Theta, \e) = \frac{\pi}{3}\,\frac{\left(2+2\e-\e^{2}\right)\!\s\sqrt{1-\s[2]}}{(1-\e)^{4}} - \frac{\pi}{2}\,\frac{\e(2-\e)\!\left(2 - \left(2-2\e+\e^{2}\right)\!\s[2]\right)\!\s}{(1-\e)^{5}\!\sqrt{1-\s[2]}}\,\eln \\
&\RepQuad{5}+ \pi\,\frac{\e^{2}(2-\e)^{2}}{(1-\e)^{5}\!\sqrt{1-\s[2]} \ssqrt}\,\sln
\end{split}
\end{align}
This correlation function is plotted in the upper right panel of Fig.~\ref{fig:redshiftAstrometryCorrelation}.

\subsection{Vectorial correlations}
\noindent
The \textsc{pta}-astrometry correlation for a background of sub-luminal vectorial \textsc{gw}s \(\G_{z\theta}^{\mathsc{X}, \mathsc{Y}} (\Theta, \e)\) is given below. The correlation function \(\G_{z\theta}^{\mathsc{Y}} (\Theta, \e)\) vanishes identically, so the correlation is given solely by the function \(\G_{z\theta}^{\mathsc{X}} (\Theta, \e)\).
\begin{align}
\begin{split}\label{eq:ptaAstrometryVectorial}
&\G_{z\theta}^{\mathsc{X}, \mathsc{Y}} (\Theta, \e) = - \frac{4\pi}{3}\,\frac{\left(5-4\e+2\e^{2}\right)\!\s\sqrt{1-\s[2]}}{(1-\e)^{4}} + 2\pi\,\frac{\left(2 - \e(2-\e)\s[2]\right)\!\s}{(1-\e)^{5}\!\sqrt{1-\s[2]}}\,\eln \\
&\RepQuad{5}- 4\pi\,\frac{\e(2-\e) + 2(1-\e)^{2}\s[2]}{(1-\e)^{5}\!\sqrt{1-\s[2]} \ssqrt}\,\sln
\end{split}
\end{align}
This correlation function is plotted in the lower left panel of Fig.~\ref{fig:redshiftAstrometryCorrelation}.

\subsection{Scalar longitudinal correlations}
\noindent
The \textsc{pta}-astrometry correlation for a background of sub-luminal scalar longitudinal \textsc{gw}s \(\G_{z\theta}^{\mathsc{L}} (\Theta, \e)\) is given below. The luminal limit (\(\e = 0\)) of this function does not exist unless the pulsar terms are included, and the correlation cannot be derived analytically.
\begin{align}
\begin{split}\label{eq:ptaAstrometryLongitudinal}
&\G_{z\theta}^{\mathsc{L}} (\Theta, \e) = \frac{2\pi}{3}\,\frac{\left(5-4\e+2\e^{2}\right)\!\s\sqrt{1-\s[2]}}{(1-\e)^{4}} - \pi\,\frac{\left(\left(3-2\e+\e^{2}\right) - \left(2-2\e+\e^{2}\right)\!\s[2]\right)\!\s}{(1-\e)^{5}\!\sqrt{1-\s[2]}}\,\eln \\
&\RepQuad{5}+ 2\pi\,\frac{1}{(1-\e)^{5}\!\sqrt{1-\s[2]} \ssqrt}\,\sln
\end{split}
\end{align}
This correlation function is plotted in the lower right panel of Fig.~\ref{fig:redshiftAstrometryCorrelation}.

\end{appendices}
\end{document}